\documentclass[aps,twocolumn,showpacs,PRR,longbibliography,superscriptaddress]{revtex4-2}
\usepackage{mathtools}
\usepackage{bm}
\usepackage{dsfont,amsthm,amsbsy}
\usepackage{verbatim}
\usepackage{amssymb}
\usepackage{amsmath}
\usepackage{bbm}
\usepackage{graphicx}
\usepackage{epstopdf}
\usepackage{subfigure}
\usepackage{natbib}
\usepackage{epsfig}
\usepackage{amsfonts}
\usepackage{mathrsfs}
\usepackage{sidecap}
\usepackage{lipsum}
\usepackage[colorlinks,linkcolor=blue,citecolor=blue,anchorcolor=blue,urlcolor=blue]{hyperref}
\usepackage{hyperref}
\usepackage{resizegather}
\usepackage{tikz}
\usepackage{float}
\usepackage{mathbbol}
\usepackage[normalem]{ulem}
\usepackage{cancel}
\usepackage{upgreek}
\usepackage[dvipsnames]{xcolor}

\newcommand{\bl}{\begin{aligned}}
\newcommand{\el}{\end{aligned}}

\def\be{\begin{equation}}
\def\ee{\end{equation}}

\def\bi{\begin{itemize}}
\def\ei{\end{itemize}}
\def\bn{\begin{enumerate}}
\def\en{\end{enumerate}}
\def\bea{\begin{eqnarray}}
\def\eea{\end{eqnarray}}
\def\no{\nonumber}
\def\ba{\begin{array}}
\def\ea{\end{array}}
\def\bd{\begin{displaymath}}
\def\ed{\end{displaymath}}

\begin{document}
\title
{Entanglement generation and scaling from noisy quenches \\ across a quantum critical point}


\author{R. Jafari}
\email[]{raadmehr.jafari@gmail.com}
\affiliation{Department of Physics, Institute for Advanced Studies in Basic Sciences (IASBS), 45137-66731 Zanjan, Iran}
\affiliation{School of Quantum Physics and Matter Science, Institute for Research in Fundamental Sciences (IPM), 19395-5531 Tehran, Iran}
\affiliation{Department of Physics, University of Gothenburg, SE 412 96 Gothenburg, Sweden}

\author{J. Naji}
\email[]{j.naji@ilam.ac.ir}
\affiliation{Department of Physics, Faculty of Science, Ilam University, Ilam, Iran}

\author{A. Langari}
\email[]{langari@sharif.edu}
\affiliation{Department of Physics, Sharif University of Technology, 11155-9161, Tehran, Iran}

\author{Vahid Karimipour}
\email[]{vahid@sharif.edu}
\affiliation{Department of Physics, Sharif University of Technology, 11155-9161, Tehran, Iran}

\author{Henrik Johannesson}
\email[]{henrik.johannesson@physics.gu.se}
\affiliation{Department of Physics, University of Gothenburg, SE 412 96 Gothenburg, Sweden}


\begin{abstract}
We study the impact of noise on the dynamics of entanglement in the transverse-field Ising chain, with the field quenched linearly across one or both of the quantum critical points of the model. 
Taking concurrence as a measure of entanglement, we find that a quench generates entanglement between nearest- and next-nearest-neighbor spins, with noise reducing the amount of entanglement. Focusing on the next-nearest-neighbor concurrence, known to exhibit Kibble-Zurek scaling with the square root of the quench rate in the noiseless case, we find a different result when noise is present: The concurrence now scales logarithmically with the quench rate, with a noise-dependent amplitude. This is also different from the ``anti-Kibble-Zurek" scaling of defect density with quench rate when noise is present, suggesting that noisy entanglement generation is largely independent of the rate of defect formation. Intriguingly, the critical time scale beyond which no entanglement is produced by a noisy quench scales as a power law with the strength of noise, with the same exponent as that which governs the optimal quench time for which defect formation is at a minimum in a standard quantum annealing scheme.
\end{abstract}

\maketitle
\section{Introduction} 

The study of quantum phase transitions (QPTs) form a core area of many-body physics. Such transitions $-$ between ground states of different
symmetry \cite{SubirBook} or topology \cite{ContinentinoBook} $-$ play out at zero temperature by varying a control parameter. As markers, physical observables exhibit scaling and universality when approaching quantum criticality, reflecting the divergence of intrinsic length and time scales.

Experimental advances in quantum simulations $-$ using ultracold atoms \cite{greiner2002}, trapped ions \cite{Islam2011}, superconducting qubits \cite{Guo2019}, 
Rydberg atoms \cite{Keesling2019}, or annealing devices \cite{Hauke2020} $-$ have opened new avenues
to studies of QPTs, particularly as it comes to their impact on nonequilibrium dynamics of many-body systems. Here, a key theoretical paradigm is that of the Kibble-Zurek mechanism (KZM)
\cite{Kibble1976,Zurek1985} extended to the quantum regime \cite{Damski2005,Zurek2005,Polkovnikov2005,Dziarmaga2005,Cherng2006,Uhlmann2007,Bando2020,Subires2022}: A system slowly driven across a continuous QPT is predicted to violate the conventional adiabaticity conditions \cite{Kato1950} and fall out of equilibrium, resulting in a formation of topological defects. The scenario is well supported in one dimension by model analyses of QPTs \cite{Cincio2007,Saito2007,Sen2008,Damski2010,Puebla2019,Rams2019} and by high-precision experimental results from quantum simulators, using an array of Rydberg atoms \cite{Keesling2019} and a D-wave quantum annealing device \cite{Bando2020}. Support for KZM is also building beyond one dimension, from theory and numerics \cite{Polkovnikov2005,Sengupta2008,Mondal2008,Chesler2015,Schmitt2022,Dziarmaga2023,Deng2024} as well as from experiments \cite{Ebadi2023}.

Central to KZM is a universal scaling law that governs how the density of defects $n_0$ scales with the rate (inverse {\em time scale}) $1/\tau$ of a ramped quench (also known as {\em linear} quench) across a continuous QPT: $n_0 \sim \tau^{-\beta}$ with $\beta = d\nu/(1+z\nu)$, where $\nu$ and $z$ are the critical correlation and dynamic exponents respectively, and $d$ is the dimensionality of the system. The subindex on $n_0$ signifies that the quench is noiseless.  

The inverse of the defect density determines the average length $\lambda$ of the ordered domains in the broken-symmetry phase. The same length  $\lambda$ appears to control also the amount of entanglement generated by the quench, suggesting a connection between defect formation and entanglement production. Evidence for this in the case of a ramped quench $-$ from paramagnetic to ferromagnetic phase of the transverse-field Ising (TFI) chain $-$ was given by Cincio {\em et al.} in Ref. \cite{Cincio2007}, confirming and extending results in Ref. \cite{Cherng2006}. Cincio {\em et al.} showed that the amount of entanglement entropy of a block with $L$ spins generated by a quench saturates at $\approx \frac{1}{6}\log_2 \sqrt{\tau}$ when $L$ is large. Knowing that the block entanglement in one dimension saturates at $\frac{1}{6}\log_2 \lambda$ with $\lambda < L$ an intrinsic length scale \cite{Calabrese2005}, this is in accord with the KZM scaling law that predicts that $\lambda$ for the quenched transverse-field Ising chain is given by $\lambda \sim \sqrt{\tau}$. The authors propose that the scenario is generic, and that the process that determines the defect density in the out-of-equilibrium dynamics of a QPT is also responsible for the entanglement that is generated. 

A different take was presented by Sengupta and Sen in Ref. \cite{Sengupta2009}. These authors probed the two-spin entanglement generated by a power-law quench across both quantum critical points $h = h_c = \pm 1$ in the XY chain with a transverse field $h$, with the quench starting [ending] deep in a paramagnetic phase, with $h_{\text{initial}} \rightarrow -\infty$ $[h_{\text{final}} \rightarrow \infty]$. It was found that when the time scale $\tau$ of the quench is larger than a critical value, $\tau > \tau_0$, only spins separated by an even number of lattice spacings get entangled, while for $\tau < \tau_0$ there is no two-spin entanglement; instead the entanglement becomes multipartite. In the first case, the concurrence that measures the two-spin entanglement scales as $\sqrt{\alpha/\tau}$, where $\alpha$ is the exponent of the power-law quench. Again, the entanglement production for a slow quench is seen to scale with the square root of the quench rate, similar to the defect density predicted by KZM for the Ising and XY chains with a transverse field. Additional results on the transverse-field XY chain, {\color{black} pertaining also to entanglement negativity and quantum discord} are in line with this result \cite{Patra2011,Nag2011}, {\color{black} with related findings in Ref. \cite{Nag2016}. Subsequent works have explored how the decoherence processes that compete with the entanglement generation scale with the quench rate \cite{Nag2012,Suzuki2016}.} 

The {\color{black}analytical and numerical} findings in Refs. \cite{Cherng2006,Cincio2007,Sengupta2009,Patra2011,Nag2011}, although limited to simple one-dimensional models, indeed suggest that a quench across a QPT, no matter how slow, will invariably lead to entanglement generation. Moreover, the KZM-type scaling of entanglement, {\color{black} as measured by concurrence, negativity, and quantum discord}, indicates that the break up of an ordered phase by defects lies at the very heart of the process. {\color{black} A simple argument that directly links the scaling of concurrence to that of defect formation is contained in the analysis in Ref. \cite{Patra2011}. Given the nonlocality of entanglement, one may have thought that the scaling of entanglement generation with $\tau$ should rather be governed by density {\em correlators} which, as shown in Refs. \cite{Roychowdhury2021,Dziarmaga2022}, cannot be derived solely from the assumptions underlying KZM, and hence go beyond the scaling of defect density formation. Importantly, however, and as shown explicitly in Ref. \cite{Roychowdhury2021} for the two-point density correlators, the relevant length and time scales that appear in these correlators {\em do} obey KZM scaling. This resolves a potentially thorny conceptual issue.} 

It is interesting to inquire about the robustness of the above scenario against perturbations. As a case in point, in any experiment where energy is transferred into or out of an otherwise isolated system via a quench, there will be time-dependent fluctuations (``noise") in the transfer, realizing an unavoidable type of perturbation. Examples include fluctuations in the effective magnetic field applied to a lattice of trapped ions \cite{Britton2012} and noise-induced heating from amplitude fluctuations of the lasers forming an optical lattice \cite{Chen2010,Doria2011}. Noise is also frequently used to model the leakage of quantum information out into an environment, with the process of decoherence simulated via an ensemble average of classical noise \cite{Budini2001,Chenu2017,Yang2017}. In short, understanding the effects of noise when quenching through a QPT is important for the modeling and design of an experiment, as well as for interpreting the results \cite{Crow2014,Ecker2019,Urbanek2021}.

The first steps in this direction were taken by Dutta {\em et al.} \cite{Anirban2016} who showed that an isolated system driven across a QPT by a noisy control field exhibits an ``anti-Kibble-Zurek" (AKZ) behavior, whereby slower driving results in a higher density of defects. The authors argued that the KZM dynamics decouples from noise-induced effects, leading to an additive scaling formula for the density of defects in the limit of weak noise, $n \sim r\tau + c\tau^{-\beta}$, with $r$ the rate at which the presence of noise generates defects, and with $c$ a nonuniversal constant. When the first term dominates, the anti-Kibble-Zurek behavior becomes manifest, as fleshed out in later theoretical \cite{Gao2017,Puebla2020,Singh2021,Singh2023,Iwamura2024,Sadeghizade2025} and experimental \cite{Ai2021} studies {\color{black}(cf. Appendix A)}.                

This raises the question about entanglement production from noisy quenches across QPTs. Does the presumed connection between production of entanglement and defects still hold? When the quench is slow, how does the generated entanglement depend on the time scale $\tau$ of the quench? And how does the intensity of the noise influence the entanglement production? 

We here address these questions in the setting of the TFI chain, the same model as used in the case studies in Refs. \cite{Anirban2016,Gao2017}. Specifically, we analyze the two-spin entanglement generated by noisy quenches across the QPTs of the model, taking white noise as a representative, and added as a classical stochastic variable to the transverse field. Since one does not have access to individual noise realizations in an experiment, we consider the ensemble average over many realizations, with the average density matrix $\langle\rho\rangle$ emerging as the relevant quantity of interest. To obtain $\langle\rho\rangle$, we solve an exact master equation for the quench dynamics averaged over the noise distribution. By numerically solving the master equation, we then extract the average entanglement produced by an ensemble of noisy quenches. When two-spin entanglement is present, as measured by concurrence, we find that it scales logarithmically with the time scale $\tau$ when the quenches are slow, with an amplitude  that depends on the intensity $\xi$ of the corresponding white-noise distribution. This is different from the behavior of the defect density in the presence of white noise as predicted by the anti-Kibble-Zurek additive formula in Refs. \cite{Anirban2016,Gao2017}. As a consequence, the notion that entanglement and defect formation are intrinsically linked becomes less persuasive. A possible counterpoint is our finding that the time scale $\tau_c(\xi)$ beyond which no two-spin entanglement is produced scales as a power law of the noise intensity with the very same exponent as that of the optimal quench time for which the density of defects is minimized in a quantum annealing scheme \cite{Anirban2016,Gao2017}. Whereas a direct, quantitative link between two-spin entanglement generation and defect formation seems unlikely in the presence of noise, this latter result hints at a possible connection, albeit more complex, between the two phenomena. 


To prepare for our analysis, in the next section we set up the formalism and define the model. As a backdrop, we then revisit the simple case of a noiseless quench in Sec. III. This allows us to present a new result: Different from a noiseless symmetric ``very-large-field" quench as discussed in Ref. \cite{Sengupta2009} (i.e., with $h_{i} = -h_{f} \rightarrow - \infty$), we find that a generic quench generates two-spin entanglement also between nearest-neighbor spins. In Sec. IV, we turn to the main focus of our work, the case of noisy quenches. We take off from the exact master equation mentioned above, display the results from its numerical solution, and extract the scaling laws for the two-spin averaged entanglement as measured by concurrence $-$ generated by slow noisy quenches across the quantum critical points of the model. Section V contains a brief summary and outlook. Mathematical details of our analysis can be found in the Appendix.  
  
\section{Model and formalism} 

We begin by writing down the Hamiltonian of the Ising chain with periodic boundary conditions and 
subject to a noiseless transverse magnetic field $h_0(t)$,
%
\bea
\label{eq:Ising}
H_0(t) =- \sum_{n=1}^{N} \big(2Js_n^x s_{n+1}^x - h_0(t)s_n^z \big).
\eea
%

Here $s_n^{x,z}=\sigma^{x,z}_n/2$, where $\sigma^{x,z}_n$ are Pauli matrices acting at sites $n$ of a one-dimensional lattice, with $\sigma^{x,z}_{N+1}=\sigma^{x,z}_1$. 
When the magnetic field is time-independent, $h_0(t)\!=\!h$, with $J$ set to unity (as will be done from now on), the ground state is ferromagnetic with $\langle s_n^x \rangle \neq 0$ for $|h| < 1$, otherwise paramagnetic with $\langle s_n^x \rangle = 0$ , the phases being separated by quantum critical points at $h=h_c=\pm 1$ \cite{Pfeuty1970}. 

Carrying out a Jordan-Wigner (JW) transformation to spinless fermion operators $c_n^\dagger, c_n$ \cite{LSM1961}, 
%
\begin{eqnarray} \label{eq:JordanWigner}
s_n^{+} &=& \prod_{j=1}^{n-1}(1-2c^\dagger_{j} c_j)c_n, \ \ \  s_n^{-} = \prod_{j=1}^{n-1}(1-2c^\dagger_{j} c_{j})c_n^\dagger, \nonumber \\
 s_n^z &=&  c_n^\dagger c_n - 1/2, 
\end{eqnarray}
%
\noindent [where $s^{\pm}_n = \sigma^{\pm}_n/2 = (\sigma^{x}_n \pm i\sigma^{y}_n)/2$], and 
introducing the discrete Fourier transforms $c_{k} = (1/\sqrt{N})\sum_n e^{ikn}c_n$ with $k = \pm (2j\!-\!1)\pi/N, j=1, 2, \ldots N/2$, 
$H_{0}(t)$ can be expressed as a sum over decoupled mode Hamiltonians $H_{0,k}(t)$,
%
\begin{equation}
\label{eq:Nambu}
H_{0}(t) = \sum_{k>0} C^{\dagger}_k H_{0,k}(t) C_k,   
\end{equation}
%
where 
%
\begin{equation} \label{eq:H0k}
H_{0,k}(t) = h_{0,k}(t)\sigma^z + \Delta_k\sigma^y.
\end{equation}
%
Here, $C_k^{\dagger} \!=\! (c_k^{\dagger}\, c_{-k})$ and $C_k \!=\! (c_k\, c_{-k}^\dagger)^T$ are Nambu spinors, 
while $h_{0,k}(t) = h_0(t)-\cos(k)$ and $\Delta_k = \sin(k)$.  
We have taken the number of sites $N$ and the fermion parity $e^{i\pi\sum_{n=1}^Na_n^\dagger a_n}$ to be even, with the assignment of momenta reflecting
the resulting antiperiodic boundary condition for the JW fermions \cite{Mbeng2024} . 
Note that the Pauli matrices that appear in $H_{0,k}(t)$ act on the Nambu spinors and should not be mixed up with
the matrices that represent spin operators in Eq. (\ref{eq:Ising}).

By projecting the Nambu spinor $C_k^\dagger$ onto the instantaneous eigenstates of $H_{0,k}(t)$, one obtains the Bogoliubov operators $\{\gamma_k^\dagger(t), \gamma_{k}(t)\}$, with 
%
\begin{eqnarray}  \label{eq:Bogoliubov}
\gamma_k^\dagger(t)& =& \cos\theta_k(t) c_k^\dagger -i\sin\theta_k(t) c_{-k} \nonumber \\
\gamma_{k}(t) &=& \cos\theta_k(t)  c_k+i\sin\theta_k(t) c_{-k}^\dagger
\end{eqnarray}
%
set by the angle $\theta_k(t) = \arctan ({\Delta_k}/{h_{0,k}(t)})/2$. In this basis, the Hamiltonian in Eq. (\ref{eq:Nambu}) takes the diagonal form
%
\begin{equation} \label{eq:diagonal}
H_0(t) = \sum_{k} \varepsilon_k(t) \gamma_k^\dagger(t) \gamma_k(t) - \sum_{k>0} \varepsilon_k(t), 
\end{equation}
%
with $\varepsilon_k(t)=\sqrt{h^{2}_{k}(t)+\Delta_k^2}$.
The eigenstates of $H_0(t)$ factorize into quasiparticle states labeled by $k$. Introducing $|0\rangle$ as the vacuum of the JW fermions $\{c_k,c_k^\dagger\}$, these states can be written as
%
\begin{eqnarray} \label{eq:Bogostates}
|\phi_k^+(t)\rangle &=& (-i\sin\theta_k(t) +\cos\theta_k(t) c_k^\dagger c_{-k}^\dagger) |0\rangle, \nonumber \\
|\phi_k^-(t)\rangle &=& (\cos\theta_k(t) -i\sin\theta_k(t) c_k^\dagger c_{-k}^\dagger) |0\rangle, 
\end{eqnarray}
%
with corresponding energies $\varepsilon^{\pm}_k(t) = \pm \varepsilon_k(t)$. When $N$ is large, 
the two states $|\phi_k^{\pm}(t)\rangle$ become near-degenerate in the limit $k \rightarrow \pi \ (k \rightarrow 0)$ when reaching 
the critical point $h_0(t) \!=\! - 1 $ ($h_0(t) \!=\! + 1 $), reflecting the avoided level crossing between the two energy bands $\{\varepsilon_k^{\pm}\}$ of the mode Hamiltonian $H_{0,k}(t)$ at criticality. 
Under a quench of time scale $\tau$, one expects a mode with wave number $k$ to remain in its instantaneous eigenstate only if 
%
\begin{equation} \label{eq:Vitanov}
\frac{1}{\tau\varepsilon_k(t)} \rightarrow 0,
\end{equation}
%
with $2\varepsilon_k(t)$ the band gap.
This limit showcases the breakdown of adiabaticity when the quench rate $1/\tau$ is finite and/or the field approaches one of the critical points of the TFI chain. 
%
\begin{figure*}[t]
\centerline{
\includegraphics[width=0.38\linewidth]{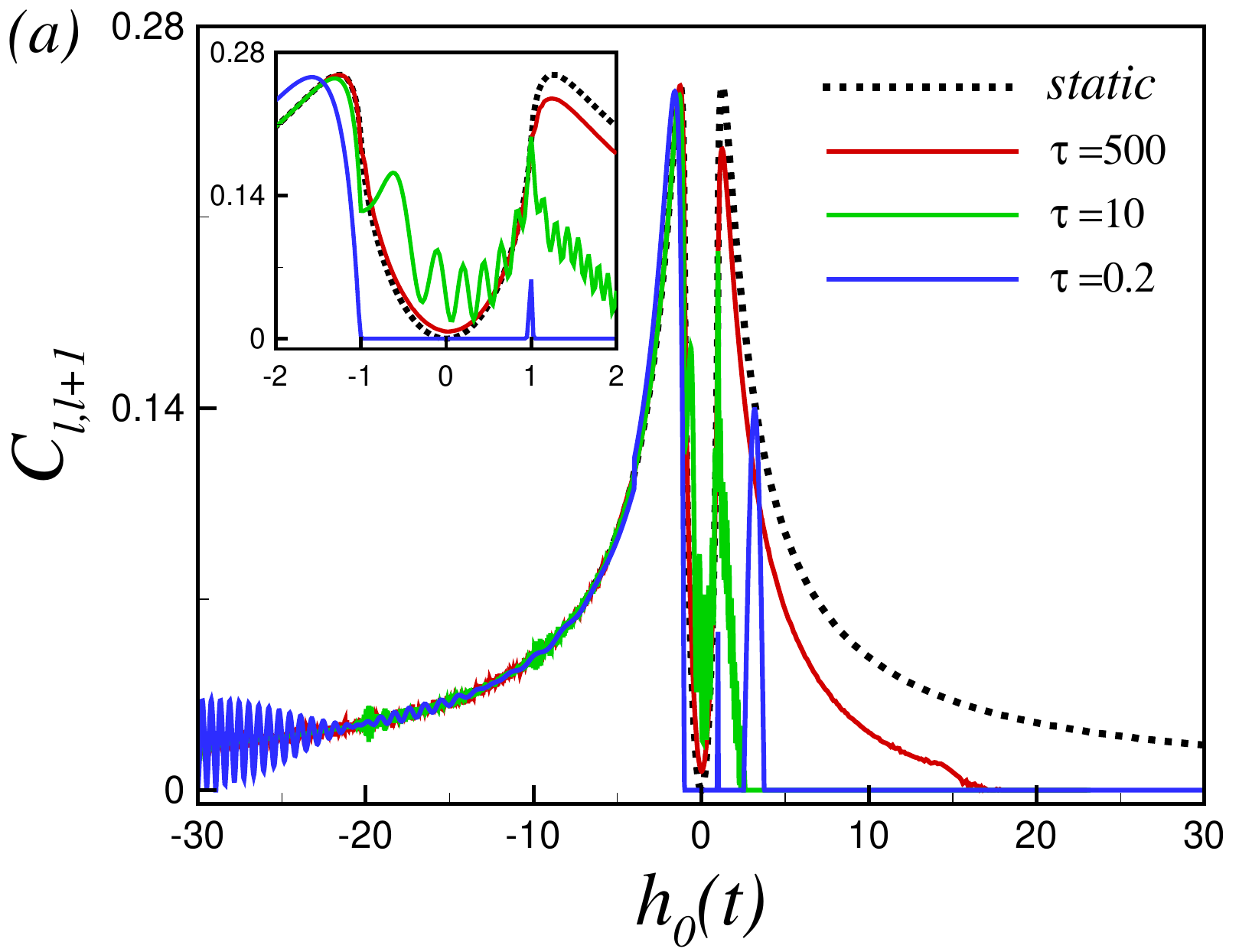}
\hspace{1.5cm}
\includegraphics[width=0.38\linewidth]{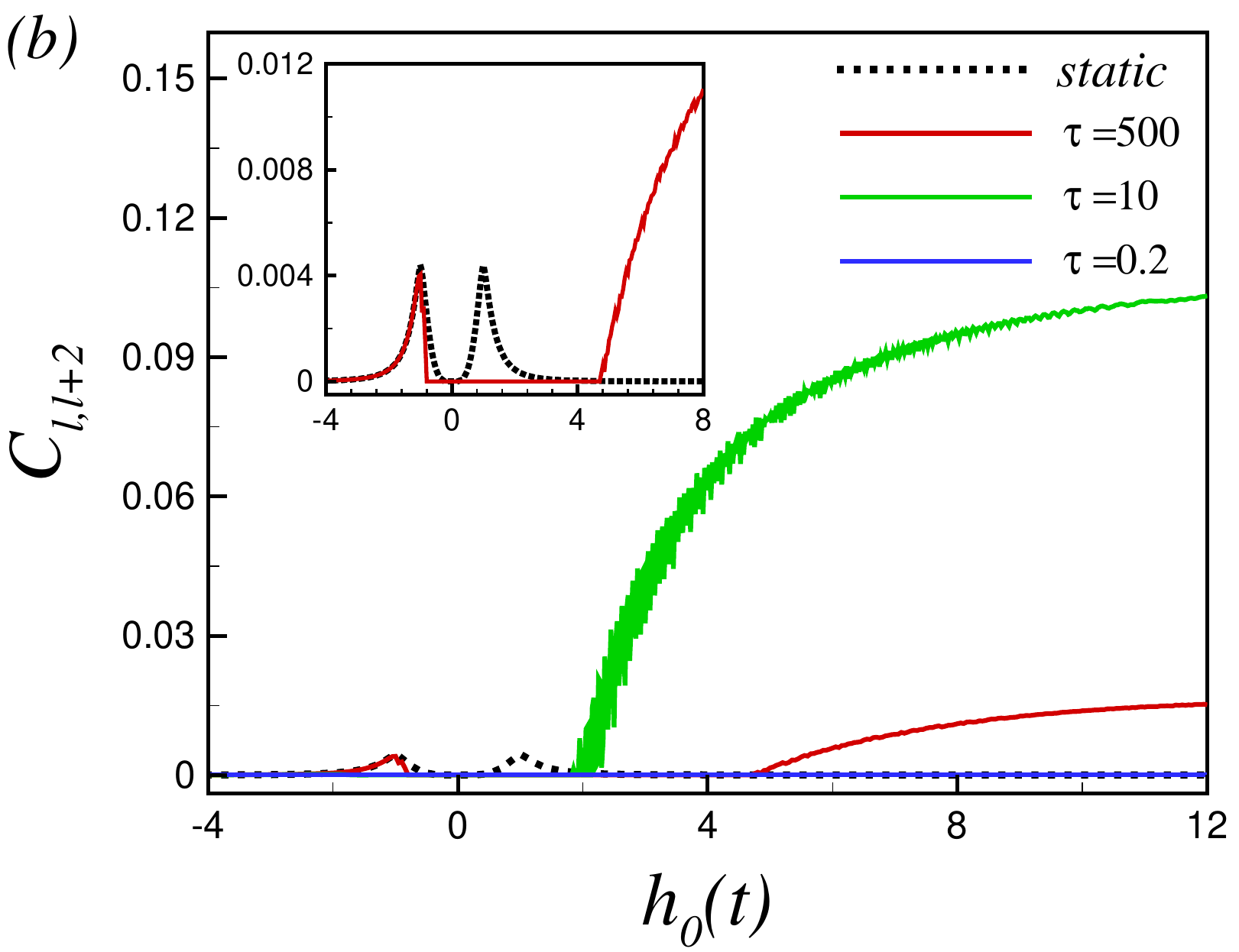}}
\caption{(Color online) Concurrence between two spins at time $t$ for a noiseless ramped quench of the transverse field in the TFI chain, from $h_i = -30$ at time $t=t_i$ to $h_f=h_0(t)$ for different values of time scale $\tau$:
(a) nearest-neighbor spins, and (b) next-nearest-neighbor spins. System size: $N=200$.}  
\label{fig1}
\end{figure*}
%

In the case of a  noiseless ramped quench we can write {\color{black} $h_0(t) \!=\!(t-t_i)/\tau +h_i$, with $\tau = (t_f-t_i)/(h_f-h_i)$ the time scale of the quench,  
where $h_i\, (h_f)$ is the initial (final) value of the ramp field at time $t_i\, (t_f)$.} To address the problem of noisy quenches, a direct approach would be to add a stochastic variable $\eta(t)$ to the ramp field, making the replacement {\color{black} $h_0(t) \rightarrow h(t)=h_0(t) + \eta(t)$} in the formulas above, compute the observable of interest $-$ or, in our case, an entanglement measure $-$ and then take an average over many noise realizations $\eta(t)$. In practice this is a cumbersome procedure, which, however, can be cut short by using an exact master equation, to be discussed in Sec. IV (see also Appendix C). 

A standard choice for this class of applications is to take the noise to be                             
Gaussian with vanishing mean, $\langle \eta({t})\rangle=0$, and with Ornstein-Uhlenbeck two-point correlations \cite{CoxMiller1965}:
%
\begin{equation}
\label{eq:noise}
\langle \eta({t})\eta({t}')\rangle=\frac{\xi^{2}}{2\tau_n}e^{-|{t}-{t}'|/\tau_n}.
\end{equation}
%
Here, $\tau_n$ is the noise correlation time and $\xi$ the noise amplitude for fixed $\tau_n$. For simplicity, and for comparison with the results in Refs. \cite{Anirban2016,Gao2017,Puebla2020,Singh2021,Singh2023,Iwamura2024,Baghran2024}, we shall confine our attention to the common white-noise limit, 
\begin{equation} \label{eq:white}
 \langle \eta({t})\eta({t}')\rangle = \xi^2 \delta(t-t'), 
\end{equation}
obtained by letting $\tau_n \rightarrow 0$ in Eq. (\ref{eq:noise}). Let us add that our assumption that the noise is uniform in space, implicit in the formulas above, is a further simplification. Future studies that model specific experimental setups may have to revoke it. For now, in our attempt to address the problem of entanglement generation from a noisy quench, we shall keep it.

\section{Entanglement generation from a noiseless ramped quench}

To set the stage for our analysis of entanglement generation from noisy quenches, in this section we revisit the simpler case of a noiseless quench. Using the notation from above, we shall consider 
a ramped quench $h_0(t)$ of the transverse field with $h_i \ll h_c = -1$, where $h_c= -1$ is the quantum critical point at which the phase changes from paramagnetic to ferromagnetic. The initial state $|\Psi(t\!=\!0)\rangle$ is prepared to be the ground state of $H_{0}(t)$. The breakdown of adiabatic dynamics due to a violation of the condition in Eq. (\ref{eq:Vitanov}) implies that the time-evolved state $|\Psi(t)\rangle$ will no longer be a ground state. Instead, the factorization of $|\Psi(t)\rangle$ into states $|\psi_k(t)\rangle$,  
%
\begin{equation} \label{eq:factorization}
|\Psi(t)\rangle = \prod_{k>0} |\psi_k(t)\rangle,
\end{equation}
%
where $|\psi_k(t)\rangle$ are superpositions of the eigenstates in Eq. (\ref{eq:Bogostates}),
%
\begin{equation}  \label{eq:superpositions}
|\psi_{k}(t) \rangle = u_{0,k}(t) |\phi^{+}_k(t)\rangle+ v_{0,k}(t) |\phi^{-}_k(t)\rangle,
\end{equation}
%
will have $u_{0,k}(t) \neq 0$ for some $k$. By normalization, $|u_{0,k}(t)|^2 + |v_{0,k}(t)|^2 =1$, with $|u_{0,k}(t)|^2$ the probability to find the $k$:th mode in the upper level $|\phi^{+}_k(t)\rangle$ with energy $\varepsilon_k$ at the end of the quench at time $t$. It follows that any finite entanglement generated after the quench will have contributions from the excited states of the system. 
 
The amplitudes $u_{0,k}(t)$ and $v_{0,k}(t)$ are easily obtained numerically from the von Neumann equation $\dot{\rho}_{0,k}(t)=-i[H_{0,k}(t),\rho_{0,k}(t)]$. Here, $\rho_{0,k}(t)$ is the density operator of mode $k$ for the noiseless ramped quench, {\color{black} with matrix elements in the diagonal basis of $H_{0,k}(t)$ given by}   
%
\begin{eqnarray} \label{eq:amplitudes}
|u_{0,k}(t)|^2&=&\!\langle \phi^{+}_{k}(t)|\rho_{0,k}(t)|\phi^{+}_{k}(t)\rangle,\nonumber \\
v_{0,k}(t)u_{0,k}^{\ast}(t)&=&\langle \phi^{-}_{k}(t)|\rho_{0,k}(t)|\phi^{+}_{k}(t)\rangle,\\
v_{0,k}^{\ast}(t)u_{0,k}(t)&=&\langle \phi^{+}_{k}(t)|\rho_{0,k}(t)|\phi^{-}_{k}(t)\rangle, \nonumber
\end{eqnarray}
%
{\color{black} with $|v_{0,k}(t)|^2 = 1 - |u_{0,k}(t)|^2$.} 

We shall focus on the nonequilibrium entanglement generated between two spins on the chain as measured by the concurrence $C_{\l,m}(t)$, with $\l$ and $m$ site indices.  
$C_{\l,m}(t)$ is defined in terms of eigenvalues of matrices built out of the reduced two-spin time-dependent density matrix $\rho_{\l,m}(t)$ \cite{Wootters1998}. 
For the TFI chain, a closed analytical expression for $\rho_{\l,m}(t)$ can be achieved via a method pioneered by Lieb {\em et al.} \cite{LSM1961}, and using a parametrization in terms of the amplitudes $u_{0,k}(t)$ and $v_{0,k}(t)$ computed from Eq. (\ref{eq:amplitudes}). For details, we refer to Appendix B.  

In Fig. \ref{fig1}(a), we present our results for the concurrence $C_{\l,\l+1}$ between two nearest-neighbor spins as a function of the endpoint quench field $h_f=h_0(t)$ at time $t$, with the quench starting at time $t_i$ with $h_i = - 30$. Three graphs are plotted, for a slow ($\tau=500$), intermediate ($\tau=10$), and fast ($\tau=0.2$) quench, all for a chain with $N=200$ sites. The black dotted curve serves as a reference, showing $C_{\l,\l+1}$ for the instantaneous ground state. As anticipated, and as shown in the inset of Fig. \ref{fig1}(a), this curve is in agreement with that known in the literature for {\color{black}$h_{0}(t) \ge 1/2$} \cite{Osborne2002,Osterloh2002}. Its minimum at $h_{0}(t)=0$ reflects the multipartite entanglement of the zero-field Greenberger-Horne-Zeilinger (GHZ) ground state for a finite ferromagnetic TFI chain, implying a suppression of the two-spin entanglement.
 
The instantaneous (``static") reference curve is seen to be well approximated by the graph for a very slow quench, $\tau=500$, for $h_0(t) \lesssim 1$. This is expected  from the adiabatic theorem \cite{Kato1950}, as a consequence of the finite size $N$ of the system: The energy gap closes as $\sim 1/N$, remaining nonzero 
for any finite $N$, so that it is always possible to reach the adiabatic limit provided $\tau$ is large enough. More precisely, it has been shown that the probability $p(\tau)$ of having an adiabatic 
evolution for system size $N$ is given by $p(\tau)=1-\exp(-2\pi^3\tau/N^2)$  \cite{Cincio2007}, so that the maximum quench rate at which the evolution remains adiabatic decays as $1/N^2$. {\color{black} As the ramp field increases, with $h_0(t) \gtrsim 1$, the $\tau=500$ curve is seen to progressively deviate from the static reference, implying that entanglement generation becomes more sensitive to the nonequilibrium dynamics for stronger fields.}   

For intermediate values of the time scale, with $\tau$ in the interval $1\lesssim\tau\lesssim100$, $C_{\l,\l+1}$ exhibits oscillations, exemplified by the green curve in the inset of Fig. \ref{fig1}(a). The oscillations, originating from the phase factors of the off-diagonal elements [$v_{0,k}(t)u_{0,k}^{\ast}(t)$, $v_{0,k}^{\ast}(t)u_{0,k}(t)$] of the density matrix $\rho_{0,k}(t)$ \cite{Cherng2006,Roychowdhury2021}, become more volatile in the lower range of the interval $1\lesssim\tau\lesssim100$, but get dampened as $h_0(t)$ increases.

As seen from the blue graph for $\tau=0.2$, a further reduction of the time scale of the quench causes $C_{\l,\l+1}$ to abruptly drop to zero after crossing the first quantum critical point at $h_0(t)=-1$. $C_{\l,\l+1}$ then suddenly shoots up when crossing the second critical point at $h_0(t)=1$, with one more spike soon thereafter before vanishing identically. This striking behavior, common to all time scales $\tau\lesssim 0.5$, shows that a very fast quench across a QPT may completely wipe out the two-spin entanglement, instead triggering multipartite entanglement [cf. the instantaneous GHZ ground state at $h_0(t)=0$] {\em or} a mutually factorized two-spin state, with the latter possibly forming an entangled composite with other spins of the system.  

%
\begin{figure*}
\centerline{
\includegraphics[width=0.38\linewidth]{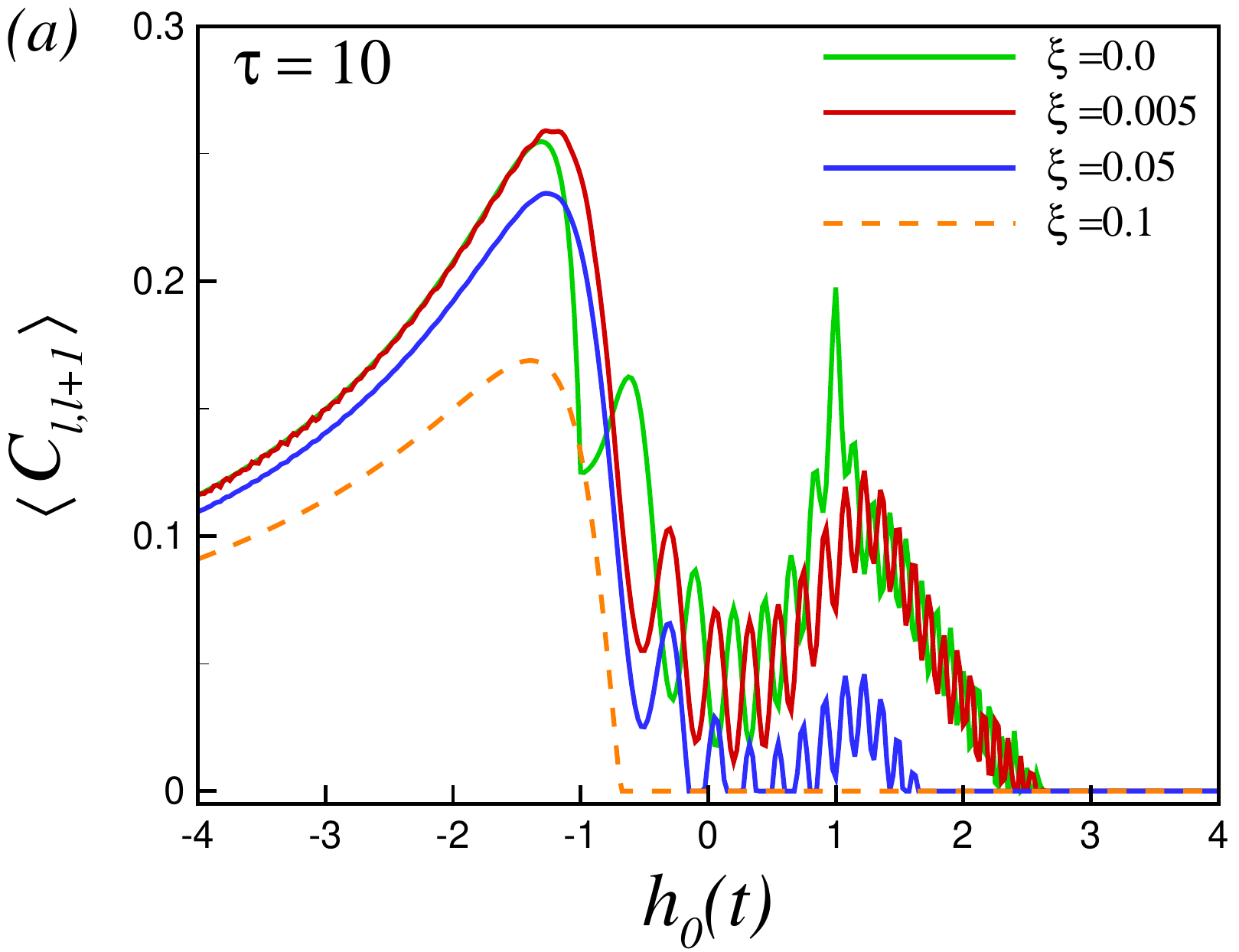}
\hspace{1.5cm}
\includegraphics[width=0.38\linewidth]{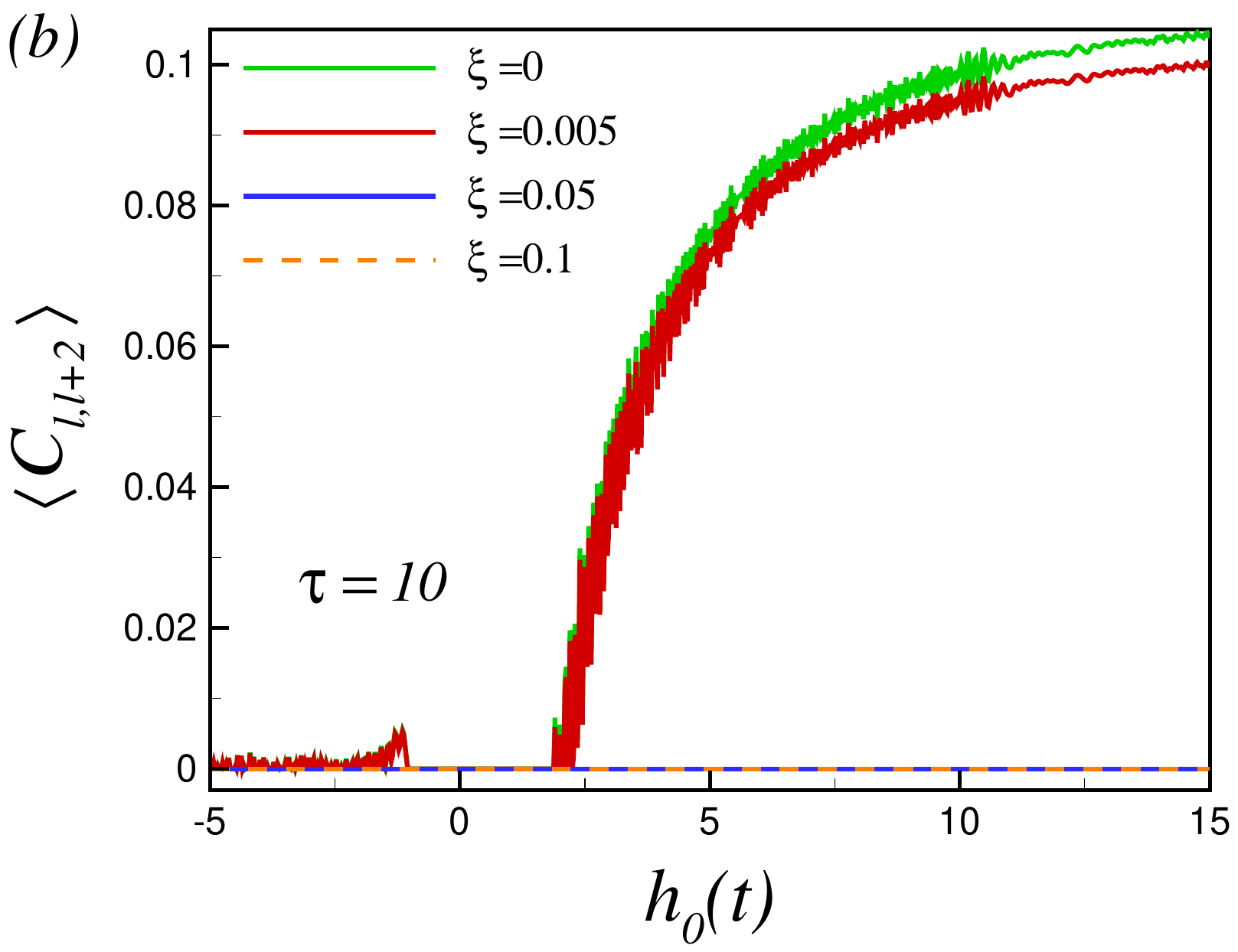}}
\caption{(Color online) Noise-averaged concurrence between two spins at time $t$ for a ramped quench of time scale $\tau=10$ with white noise of different intensities $\xi$ added to the transverse field of the TFI chain, with quench interval from $h_i=-30$ at $t=t_i$ to $h_f=h_0(t)$: (a) nearest-neighbor spins and (b) next-nearest-neighbor spins. System size: \!$N=200$.}  
\label{fig2}
\end{figure*}
%
%
\begin{figure*}
\centerline{
\includegraphics[width=0.38\linewidth]{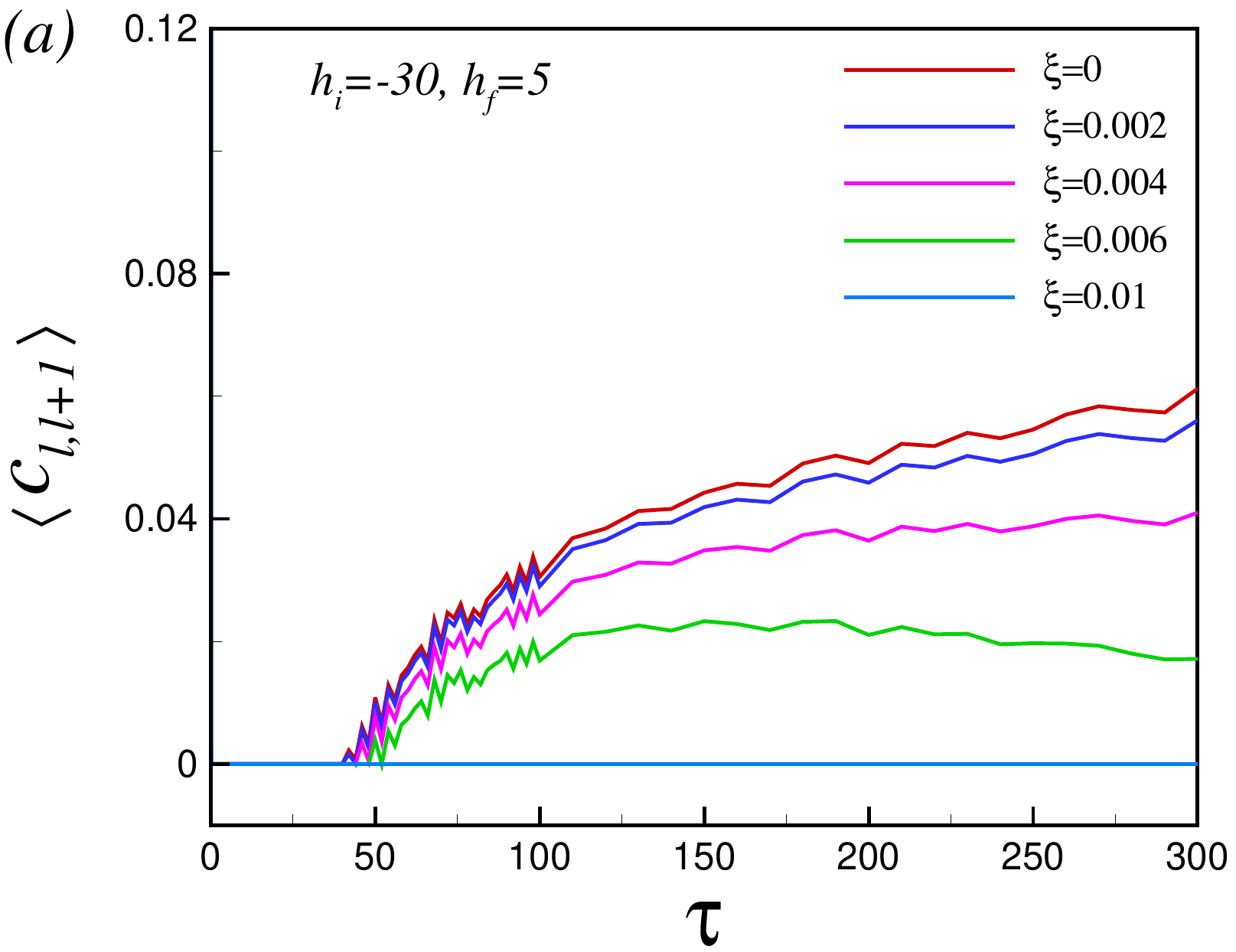}
\hspace{1.5cm}
\includegraphics[width=0.38\linewidth]{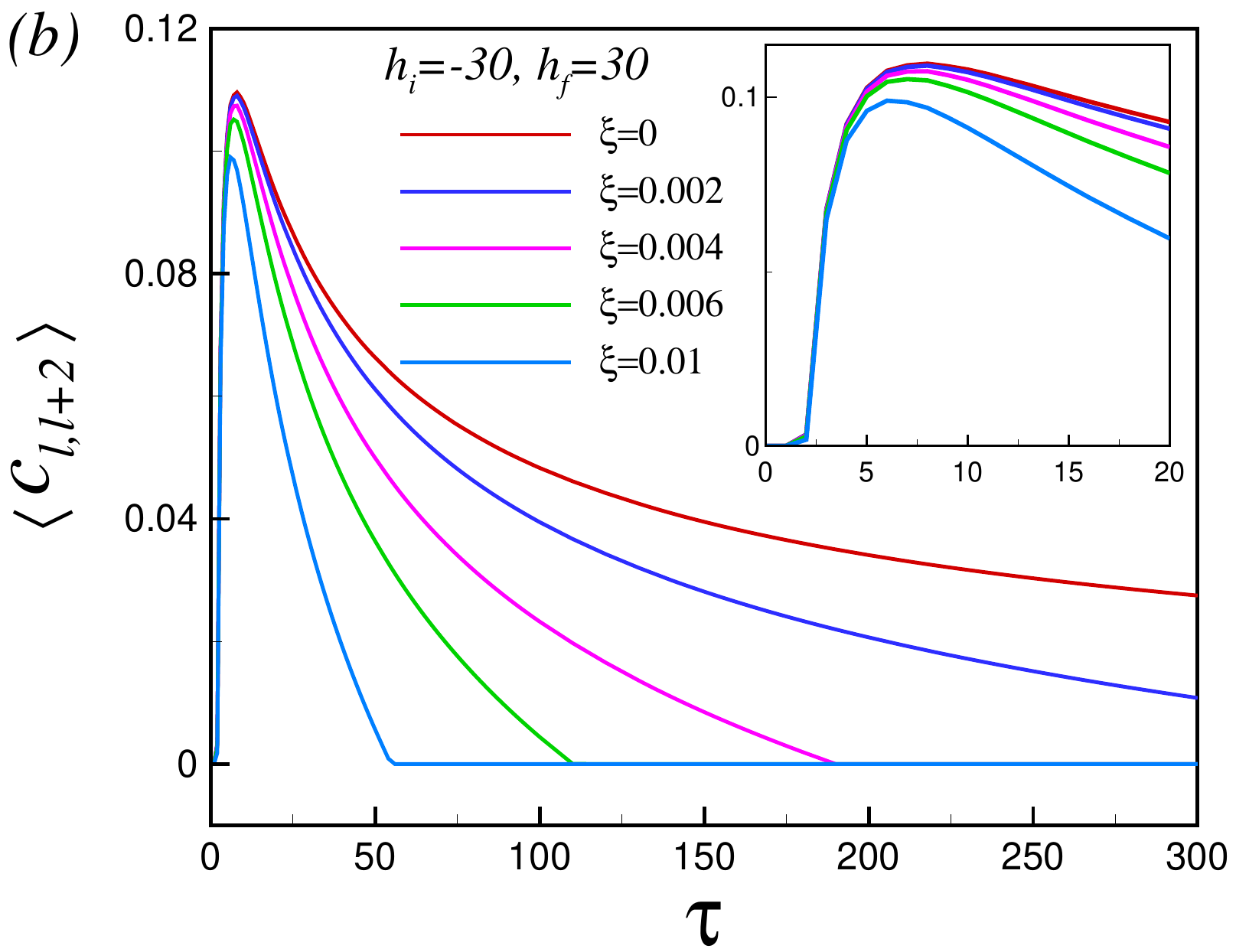}}
\caption{(Color online) Noise-averaged concurrence between two spins as a function of time scale $\tau$ for a ramped quench with added white noise of different intensities $\xi$, with quench interval from $h_i=-30$ to
(a) $h_f = 5$ for nearest-neighbor spins, and (b) $h_f = 30$ for next-nearest-neighbor spins. System size: \!$N=200$.} 
\label{fig3}
\end{figure*}
%

Independent of the choice of time scale $\tau$, all quenches generate entanglement between neighboring spins that decays to zero when $h_0(t) \rightarrow \infty$, consistent with the analytical prediction by Sengupta and Sen \cite{Sengupta2009}. These authors predicted a different type of behavior for the next-nearest-neighbor concurrence $C_{\l,\l+2}$, with two distinct asymptotics for very large $h_0(t)$: $C_{\l,\l+2} \neq 0$ [$C_{\l,\l+2} = 0$] when $\tau > \tau_0 \approx 1.96$ [$\tau < \tau_0$]. Our results for $C_{\l,\l+2}$, displayed in Fig. \ref{fig1}(b) as a function of the endpoint quench field $h_0(t)$ at time $t$, with the quench starting at time $t_i$ with $h_i = - 30$, agrees with this prediction: By extrapolating to large $h_0(t)$, we find numerically that $1.9 < \tau_0 < 2.0$.  

While consistent with the prediction of the time scale $\tau_0$ \cite{Sengupta2009}, the overall behavior of $C_{\l,\l+2}$ as shown in Fig. \ref{fig1}(b) is unexpected from the perspective of KZM. Only for slow quenches [cf. the red curve with $\tau=500$ in Fig \ref{fig1}(b)] is there a weak imprint on $C_{\l,\l+2}$ that the quench has approached or crossed the critical point $h_c = -1$ (but not $h_c=1$). Besides this, values of $C_{\l,\l+2}$ remain very small $-$ not visible in Fig. \ref{fig1}(b) $-$ for all  $h_0(t) \lesssim 2$. This contrasts with the nearest-neighbor concurrence $C_{\l,\l+1}$ which exhibits a dramatic increase when approaching the first critical point $h_c=-1$, supporting the conjecture that the entanglement generated reflects the KZM formation of defects when the dynamics turns nonadiabatic \cite{Cincio2007,Cherng2006}. One may naively have thought that the fast drop of $C_{\l,\l+1}$ in the ferromagnetic phase $-1 < h_0(t) < 1$ [cf. Fig. \ref{fig1}(a)] would influence $C_{\l,\l+2}$ by a restructuring of the entanglement with closest further neighbor spins.  But not so. Instead, the next-nearest-neighbor concurrence is driven by the postcritical dynamics when the magnitude of the transverse field becomes sufficiently large.
{\color{black}It is interesting to note that for near-adiabatic evolution (very slow quench: $\tau=500$), $C_{\l,\l+2}$ closely follows its instantaneous counterpart (static reference curve) only when $h_0(t)<h_c=-1$, as illustrated in Fig. \ref{fig1}(b). Contrary to expectations, its behavior is distinct from the reference curve for $h_0(t)>-1$. This finding, similar to that above for $C_{\l,\l+1}$ with $h_0(t) \gtrsim 1$, reflects how entanglement generation in non-equilibrium systems can differ from its equilibrium counterpart, even during a nearly adiabatic evolution.}

\section{Entanglement generation from noisy ramped quenches} 

Let us now turn our attention to the effect from noise on entanglement generation when quenching across a critical point of the TFI chain. As discussed in Sec. I, the experimentally relevant protocol is to
repeat the quench many times, then taking an average over the outcome, in effect averaging over the noise distribution. 
Computationally, a ``brute-force" option is to repeatedly solve the stochastic Schr\"odinger equation for a large number of single noise realizations, then forming an average \cite{Jafari2024}. A simpler, more elegant way is to solve an exact noise master equation \cite{luczka1991quantum,Budini2001,Chenu2017,Kiely2021} for the noise-averaged density matrix $\rho_k(t) = \langle \rho_{\eta,k}(t) \rangle$,
with $\rho_{\eta,k}(t)$ the density matrix for a given mode $k$ and white-noise realization $\eta(t)$ [Eq. (\ref{eq:white})].
 Explicitly,
%
\begin{eqnarray} \label{eq:AddMaster}
\no
\dot{\rho}_{k}(t)=-i[H_{0,k}(t),\rho_{k}(t)]-\frac{\xi^2}{2}[\sigma^z , [\sigma^z,\rho_{k}(t)]],\\
\label{eq:master}
\end{eqnarray}
%
with a derivation in Appendix C. The master equation has the form of a von Neumann equation with an added term $-({\xi^2}/{2})[\sigma^z , [\sigma^z,\rho_{k}(t)]]$ representing the effect from the noise. By numerically solving Eq. (\ref{eq:master}) and proceeding analogous to the noiseless case sketched in Sec. III {\color{black} but now using a mixed ensemble when calculating correlation functions,} we obtain the mean concurrence $\langle C_{\l,m} \rangle$ between spins at lattice sites $\l$ and $m$ as an ensemble average over the white-noise distribution $\{ \eta \}$ drawn from Eq. (\ref{eq:white}). For details, see Appendix B.

The noise-averaged nearest-neighbor and next-nearest-neighbor concurrence $\langle C_{\l,\l+1} \rangle$ and $\langle C_{\l,\l+2} \rangle $, respectively, are displayed in Fig. \ref{fig2} as a function of $h_0(t)$ for $\tau=10$ and three different values of noise intensity $\xi$ {\color{black} and noiseless case $\xi=0$ corresponding to Fig. \ref{fig1}}. With the exception of a very weak intensity, $\xi = 0.005$ in Fig. \ref{fig2}(a), the noise is seen to reduce the entanglement for almost all values of $h_0(t)$. This is expected since noise generically acts as a source of decoherence.  

To estimate the amount of entanglement reduction due to noise, we have plotted the noise-averaged concurrence in Fig. \ref{fig3} as a function of $\tau$ for four different noise intensities $\xi$, for quenches from $h_i=-30$ to $h_f=5$ $[h_f=30]$ for $\langle C_{\l,\l+1} \rangle$ $[\langle C_{\l,\l+2} \rangle]$.  {\color{black} As shown above, the concurrence between spins separated by an odd number of lattice spacings get entangled for finite $h_f$. To make this feature visible in the presence of noise, we have chosen a smaller value of $h_f$ in Fig. \ref{fig3}(a) [as compared to Fig. \ref{fig3}(b)] since $\langle C_{\l,\l+1} \rangle$ vanishes for larger values.}  
The result displayed in Fig. \ref{fig3} is qualitatively anticipated: The larger the noise intensity, the larger the entanglement reduction. 

%
\begin{figure*}[t]
\centerline{
\includegraphics[width=0.38\linewidth]{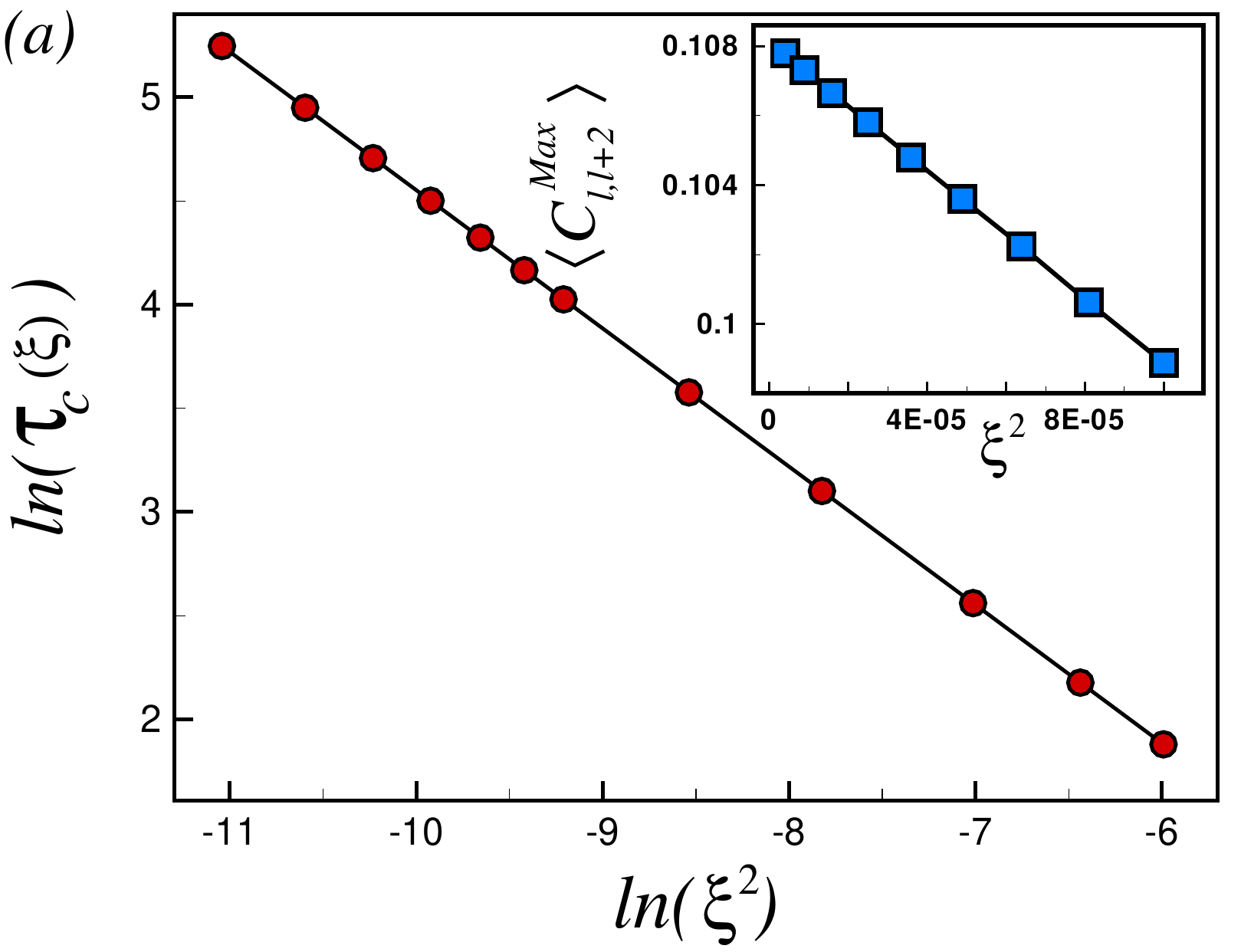}
\hspace{1.5cm}
\includegraphics[width=0.38\linewidth]{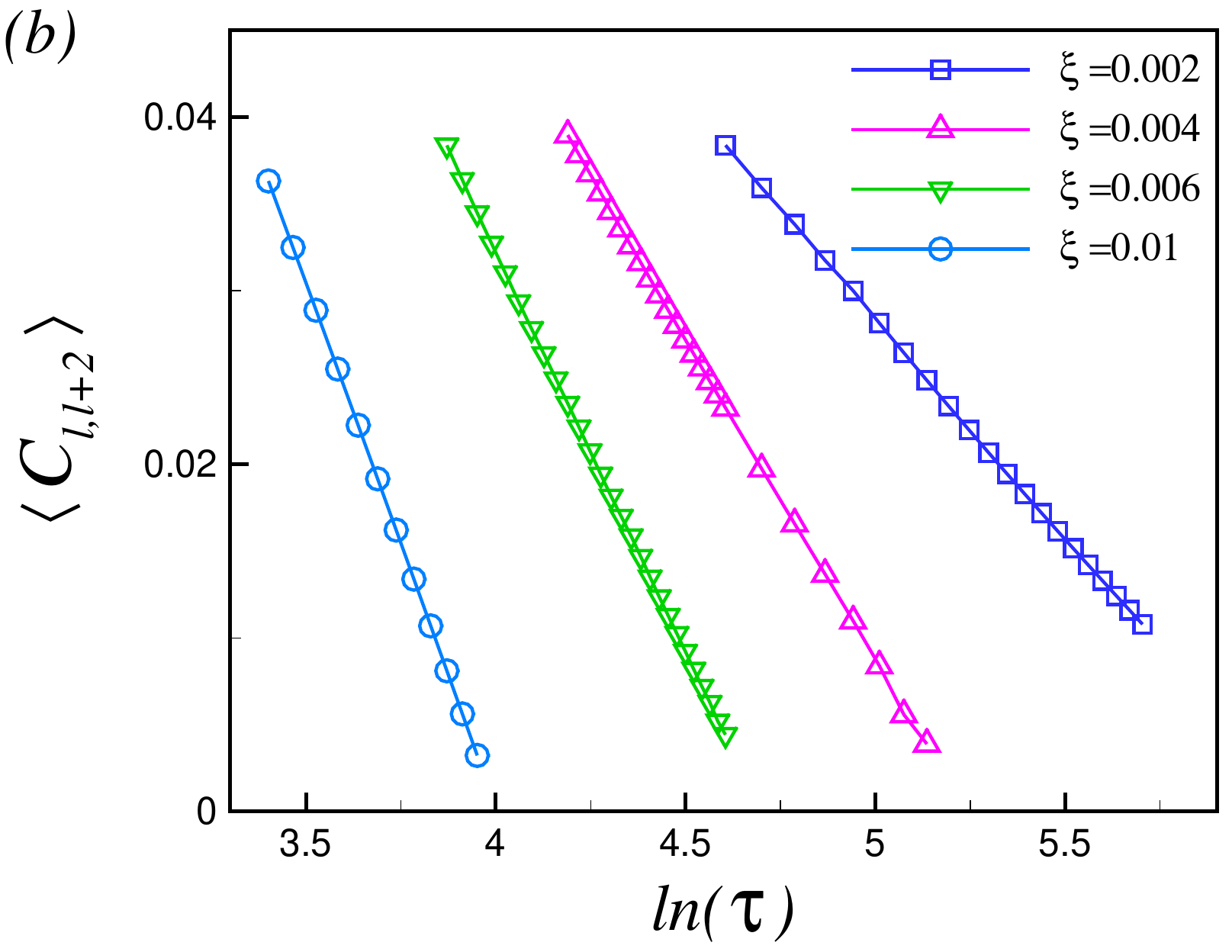}}
\caption{(Color online)  
(a) Power-law scaling $\tau_c(\xi) \sim \xi^{-\delta}$ with exponent $\delta = 0.666\pm0.001\approx2/3$, where $\xi$ is the noise intensity and $\tau_c(\xi)$ is 
the critical time scale at which the noise-averaged concurrence $\langle C_{\l,\l+2} \rangle$ vanishes. The inset shows the linear scaling of the maximum next-nearest-neighbor concurrence with
$\xi^2$. (b) Noise-averaged concurrence   
between next-nearest-neighbor spins as functions of $\ln(\tau)$
for a ramped quench with added white noise for different values of 
noise intensity $\xi$, with quench interval from $h_i=-30$ to $h_f=30$.} 
\label{fig4}
\end{figure*}
%

It is interesting to contrast our result for $\langle C_{\l,\l+2} \rangle$ with the
noiseless case discussed by Sengupta and Sen \cite{Sengupta2009}, where concurrence between spins separated by an even number of lattice bonds is analyzed. Whereas the noiseless concurrence $C_{\l,\l+2}$ in Fig. \ref{fig3}(b) vanishes when the time scale of the ramped quench goes to infinity, in agreement with Ref. \cite{Sengupta2009}, the noise-averaged concurrence $\langle C_{\l,\l+2} \rangle$ vanishes already at finite values of the time scale. Importantly, note that the entanglement reduction for fixed noise intensity increases with increasing $\tau$, indicating that for larger values of $\tau$ there is more time for noise to become effective and, by that, cause disentanglement. An analysis reveals that noise with intensity $\xi \ge 0.05$ wipes out $\langle C_{\l,\l+2} \rangle$ for all $\tau$ [cf. Fig. 2(b) when $\tau=10$], while for $0 < \xi <  0.05$ there is a critical time scale, call it $\tau_c(\xi)$, for which the spins remain entangled when $\tau_0 < \tau < \tau_c(\xi)$ (with $\tau_0 \approx1.96$ defined above).

The dependence of the critical noisy time scale $\tau_c(\xi)$ on $\xi$ is displayed in a log plot in Fig. \ref{fig4}(a), from which one extracts
\begin{equation} \label{eq:noisytimescale}
\tau_c(\xi) \sim (\xi^2)^{-\delta},
\end{equation}
with $\delta = 0.666\pm0.001\approx 2/3$. 
Comparison with $\tau_c(\xi)$ for $\xi= 0.004, 0.006$, and $0.01$ as seen in Fig. \ref{fig3}(b), implies that $\tau_c(\xi) \approx 0.12 \,(\xi^2)^{-2/3}.$ We have chosen to present the scaling in Eq. (\ref{eq:noisytimescale})
using the squared noise intensity $\xi^2$ that determines the strength of the white-noise correlations [Eq. (\ref{eq:white})]. This brings out an intriguing connection with the anti-Kibble-Zurek behavior as discussed in Ref. \cite{Anirban2016}: These authors found that the optimal quench time that minimizes the production of defects scales with $1/\xi^2$ as a power law with exponent $\delta=2/3$, the same exponent as in our Eq. (\ref{eq:noisytimescale}). While our quench protocol is different from that in Ref. \cite{Anirban2016} $-$ with the quench crossing both critical points in the TFI chain instead of just one $-$ the very same protocol as ours, applied to the XY chain (of which the TFI chain is a special case), was later found to also yield the same exponent $\delta=2/3$ \cite{Gao2017}. Whether the identity of the exponents, both found in quenches that emulate standard quantum annealing schemes, is only coincidental or signals a hidden link between the two phenomena $-$ minimization of KZM defects and vanishing of next-nearest neighbor concurrence in the presence of noise $-$ remains an open question.

Making a log plot (not shown) of the red curve for the noiseless case in Fig. \ref{fig3}(b), we immediately verify the KZM scaling law $C_{\l,\l+2}  \sim 1/\sqrt{\tau}$ established in Ref. \cite{Sengupta2009} for slow quenches, $\tau \gtrsim 150$. As displayed in Fig. \ref{fig4}(b), the curves from Fig. \ref{fig3}(b) with nonzero noise instead scale logarithmically with $\tau$ before vanishing at $\tau_c(\xi)$,
%
\begin{equation} 
\label{eq:noisescaling}
\langle C_{\l,\l+2} \rangle_{\xi} = c_\xi \ln{(\tau/\tau_c(\xi))}, 
\end{equation}
%
where $c_\xi$ is a negative noise-dependent amplitude.
We have inserted a subindex $\xi$ to explicitly specify the corresponding noise distribution from Eq. (\ref{eq:white}). 
Sampling a large number of noise-averaged concurrences $\langle C_{\l,\l+2} \rangle_{\xi}$ with $\xi \in [0.001, 0.05]$, and with the same restriction
on time scale as in Eq. (\ref{eq:noisescaling}), we find that the amplitude $c_\xi$ depends on the noise intensity as
\begin{equation} 
\label{eq:noiseexponent}
c_{\xi} = a_\xi \pm 0.001.
\end{equation}
with $a_\xi=4.100$ for $0<\xi\leqslant0.004$ and $a_\xi=3.260$ for $0.004<\xi\leqslant0.05$. It is tempting to interpret the sharp amplitude jump at $\xi = 0.004$ as signaling some kind of critical feature in the entanglement dynamics. Its elucidation also here calls for more work.

The result in Eq. (\ref{eq:noisescaling}) differs from the anti-Kibble-Zurek scaling of defect formation first found  in Ref. \cite{Anirban2016} and obtained for the same noisy quenches \cite{Gao2017} as explored by us. The authors of Refs. \cite{Anirban2016,Gao2017} showed that noise contributes a term $ \sim \tau$, added to the usual noiseless KZM scaling term $\sim 1/\sqrt{\tau}$. Whereas our power-law scaling for the vanishing of the next-nearest-neighbor concurrence in the presence of noise [Eq. (\ref{eq:noisytimescale})] may hint at a link to defect formation, the difference of the logarithmic scaling in Eq. (\ref{eq:noisescaling}) from that in Refs. \cite{Anirban2016,Gao2017} appears to rule out that the two phenomena are quantitatively connected.

{\color{black} It is important to note that the oscillations in $\langle C_{l,l+1} \rangle$ for finite values of $h_f$ [cf. Fig. \ref{fig3}(a)] preclude a direct scaling analysis similar to that for $\langle C_{l,l+2} \rangle$. This is the reason why we do not present scaling results for $\langle C_{l,l+1} \rangle$.}     

\section{Summary and outlook} 
Collecting our results for slow noisy quenches across the critical points of the TFI chain, we are led to infer that the direct link between KZM defect formation and entanglement generation suggested by the results in Refs. \cite{Cherng2006,Cincio2007,Sengupta2009,Patra2011,Nag2011} gets broken by noise. Specifically, the known KZM scaling $\sim 1/\sqrt{\tau}$ of the next-nearest neighbor concurrence \cite{Sengupta2009}, with $\tau$ the time scale of the quench, does not apply when noise is present. Instead, we find a logarithmic scaling $\sim \ln(1/\tau)$. Additional evidence that entanglement generation is nontrivially connected $-$ if at all connected $-$ to the defect formation comes from a comparison with the AKZ results by Dutta {\em et al.} \cite{Anirban2016} and Gao {\em at al.} \cite{Gao2017}. These authors found that the scaling of the defect density in presence of noise is governed by two independent terms: one noisy term $\sim \tau$, and one KZM term $\sim 1/\sqrt{\tau}$, again different from the logarithm uncovered by our work. 

The picture gets intricate by another finding of ours: The time scale at which the next-nearest neighbor concurrence vanishes scales with the squared noise intensity as $(\xi^2)^{-\delta}$, where $\delta = 0.666 \pm 0.001 \approx 2/3$. This exponent is the same as that which governs the scaling of the optimal time for minimal defect production in a standard quantum annealing scheme \cite{Anirban2016,Gao2017}. Is this a coincidence, or is it a sign that defect formation and entanglement generation are somehow linked? If the latter, our finding of the anomalous logarithmic scaling when noise is present still suggests that the two phenomena cannot be as closely intertwined as for noiseless quenches \cite{Cherng2006,Cincio2007,Sengupta2009,Patra2011,Nag2011}.

{\color{black} A related line of research, pioneered recently by one of us \cite{Asadian2025}, addresses how noise amplifies decoherence in nonequilibrium critical dynamics. The interplay between decoherence and entanglement generation in noisy critical quenches is another intriguing question that warrants further work.}

Considering the rapid progress in quantum simulations, our findings may soon be within reach of experimental tests. Quantum annealing devices have excelled in checking the KZM scaling of defect formation \cite{Bando2020}. However, while intrinsically noisy, the type of noise and its intensity are presently difficult to control on these types of platforms \cite{Albash2018}. An alternative approach, allowing for high-precision tuning of narrow-band white noise, was pioneered by Ai {\em et al.} in their experimental study of anti-Kibble-Zurek scaling of defect formation \cite{Ai2021}. These authors used a ``mixing wave" microwave device applied to a single trapped $^{171}\mbox{Yb}^+$ ion to simulate the independent mode Hamiltonians of the XY chain with a noisy transverse field (cf. our mode Hamiltonian for the TFI chain in Sec. II, being a special case of that for the XY chain). The setup proved to be well adapted to test and verify the scaling predicted by Gao {\em et al.} \cite{Gao2017}. In our present study one is interested in the scaling of {\em concurrence} with the time scale of the quench, to be recorded for different noise intensities. To obtain noise-averaged concurrence experimentally requires repeated measurements of single-spin expectation values {\em and} of appropriate spin correlation functions \cite{Laurell2024}. This requires a setup beyond that of a single trapped ion, different from the experiment in Ref. \cite{Ai2021}. While challenging, several of the experimental ingredients are already in place, holding promise for a test of our results in the near future. 

\section*{Acknowledgement}

This work is based upon research funded by Iran National Science Foundation (INSF) under Project No. 4024561.


\section*{DATA AVAILABILITY}

The data that support the findings of this article are openly available \cite{WNCC}.

\section*{Appendix A: Defect generation under noisy ramp fields}

\begin{figure}[t]
\centerline{
\includegraphics[width=0.8\columnwidth]{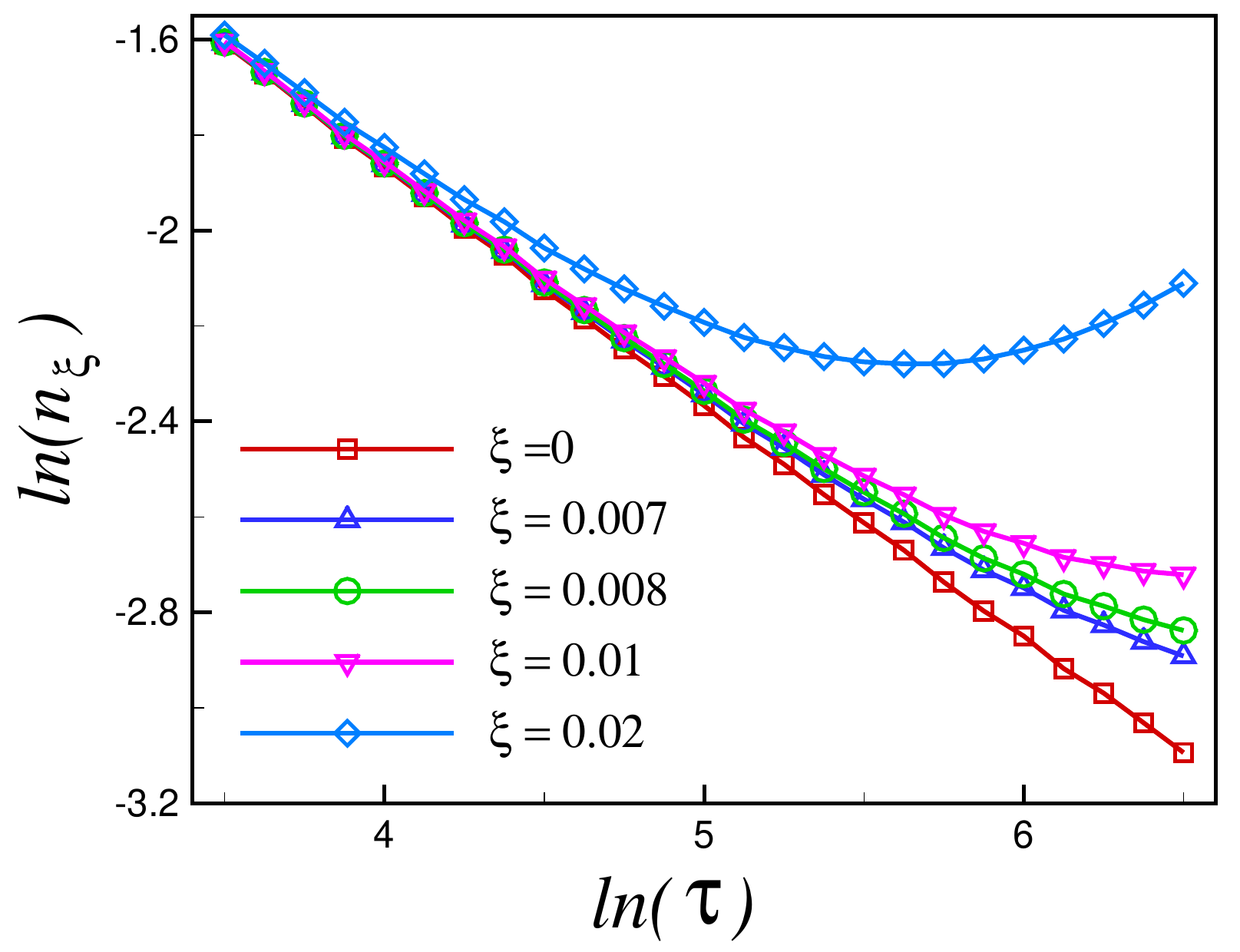}}
\caption{(Color online) {\color{black} The logarithm of the density of defects $n_{\xi}$ vs time scale $\tau$ of a ramped quench crossing a quantum critical point in the transverse field Ising chain for different noise intensities $\xi$.
The behavior of the density of defects for different values of $\tau$ and $\xi$ illustrates anti-Kibble-Zurek behavior in the presence of a noisy ramp field.}} 
\label{fig5}
\end{figure}

{\color{black}
As discussed in Sec. 1, defect generation under noisy ramp fields exhibits an AKZ behavior when crossing a quantum critical point, whereby slower driving results
in a higher density of defects \cite{Anirban2016,Gao2017,Puebla2020,Singh2021,Singh2023,Iwamura2024,Sadeghizade2025}. 

The dependence of the density of defects $n_{\xi}$ on the time scale $\tau$ of the ramp is shown in Fig. \ref{fig5} for several values of
noise intensity $\xi$. For $\xi=0$ case, numerical simulations verify the KZM scaling law $n_0 \propto \tau^{-1/2}$. As is also seen in the figure, for $\xi \neq 0$  
the effect of noise on the defect density is insignificant for small noise amplitudes $\xi$ when $\tau$ takes small and intermediate values. Hence, in these cases, the KZM prediction $n_{\xi} \propto \tau^{-1/2}$ still holds approximately.  

As the ramp time scale increases, noise-induced defects dominate the nonadiabatic dynamics, leading to a growth of the defect density with $\tau$. 
This is indicative of the AKZ regime, where enhancing the ramp time scale results in a higher density of defects in the system. In the
limit of a very large values of $\tau$, $n_{\xi}$ is completely governed by the AKZ contribution. In summary, the defect generation is controlled by two 
competing mechanisms: (1) non-adiabatic defect density, {\em suppressed} by increasing the time scale $\tau$ of the ramp; and (2) noise-induced defect density, {\em amplified} by increasing $\tau$. 
As shown in Refs. \cite{Anirban2016,Gao2017}, the interplay of these two competing influences yields a minimum of the density of defects ($n^{\text{min}}_{\xi}$) at an optimum annealing
time $\tau_{\text{opt}}$ that scales as $\tau_{\text{opt}} \propto \tau^{-\delta}$, with $\delta=0.666\pm0.001\simeq2/3$.

It is notable that the AKZ regime appears only when traversing a {\em quantum critical point} with sizable noise in the ramp field. In contrast, KZM scaling is expected to be insensitive to thermal noise in classical phase transitions: As found in Ref. \cite{Das2012}, KZM scaling for local defect densities and winding numbers hold even in the presence of thermal noise. 
}

\section*{Appendix B: Computing concurrence for time-evolved states of the TFI chain}

\setcounter{equation}{0}
\renewcommand\theequation{B\arabic{equation}}

As measure of the entanglement at time $t$ between two spins at sites $\l$ and $m$ in the TFI chain we use {\em concurrence}, here denoted $C_{\l,m}(t)$ and defined as \cite{Wootters1998}
%
\begin{equation} \label{eq:Cdef}
C_{\l,m}(t) =  \mbox{max}(0, \lambda_1(t)\! -\! \lambda_2(t)\! -\! \lambda_3(t) \!-\! \lambda_4(t)).
\end{equation}
%
The $\lambda_{i}$'s are eigenvalues, in descending order, of the product matrix  
%
\begin{equation} \label{eq:Rmatrix}
R_{\l,m}(t) = \sqrt{\sqrt{\rho_{\l,m}(t)}\tilde{\rho}_{\l,m}(t)\sqrt{\rho_{\l,m}(t)}},
\end{equation}
%
with $\rho_{\l,m}(t)$ the time-evolved reduced density matrix of the two spins, $\tilde{\rho}_{\l,m}(t)$ being the
corresponding spin-flipped matrix, $\tilde{\rho}_{\l,m}(t) = (\sigma^y_{\l}\otimes\sigma^y_m)\rho^\ast_{\l,m}(\sigma^y_{\l}\otimes\sigma^y_m)$.  
%
%
The density matrix $\rho_{l,m}$ (with suppressed time argument) is known for the TFI chain \cite{Barouch1971a,Syljuasen2003}, and can be written
%
\bea
\label{eq:APB1}
\rho_{l,m}=\left(\begin{array}{cccc}
 \kappa_{11} &0 &0 & \kappa_{14} \\
 0 & \kappa_{22} & \kappa_{23} &0 \\
 0 & \kappa_{23}^{\ast} & \kappa_{33} &0 \\
  \kappa_{14}^{\ast} &0 &0 &\kappa_{44}\,  \\
\end{array}\right),
\eea
%
with matrix elements expressed in terms of one- and (equal-time) two-point spin correlation functions,
%
\bea
\label{eq:APB2}
\no
\kappa_{11} &=& \langle s^z_{l}\rangle+\langle s^{z}_{l} s^{z}_{m} \rangle+\frac{1}{4},\\
\no
\kappa_{22} &=& \kappa_{33}=-\langle s^{z}_{l} s^{z}_{m} \rangle+\frac{1}{4},\\
\kappa_{44} &=& - \langle s^z_{l}\rangle+\langle s^{z}_{l} s^{z}_{m} \rangle+\frac{1}{4},\\
\no
\kappa_{23} &=& \langle s^{x}_{l} s^{x}_{m} \rangle+\langle s^{y}_{l} s^{y}_{m} \rangle+i(\langle s^{x}_{l} s^{y}_{m}\rangle-\langle s^{y}_{l} s^{x}_{m}\rangle),\\
\no
\kappa_{14} &=& \langle s^{x}_{l} s^{x}_{m} \rangle-\langle s^{y}_{l} s^{y}_{m} \rangle-i(\langle s^{x}_{l} s^{y}_{m}\rangle+\langle s^{y}_{l} s^{x}_{m}\rangle),
\eea
%
where $s_{\l}=\sigma^\alpha_{\l}/2, \alpha=x,y,z$ are the spin components at site $\l$ of the chain, with $\sigma_{\l}^\alpha$ the Pauli matrices.
The expectation values $\langle \ldots \rangle$ are here taken in the time-evolved spin states. 
It should be mentioned that while in a time-independent magnetic field, $h_0(t)=h$, the two-point spin correlations 
$\langle s_n^xs_m^y\rangle$ and $\langle s_n^ys_m^x\rangle$ vanish, they may take finite values when time dependent \cite{Barouch1971b}.

Singling out a reference spin for notational transparency, say at site $\l=1$, and using the JW transformation in Eq. (\ref{eq:JordanWigner}), the two-point functions in Eq. (\ref{eq:APB2}) can be expressed 
in terms of multipoint correlators of the JW fermions,
%
\bea
\label{eq:APB3}
\no
\langle s_1^xs_{1+ r}^x \rangle  &\!=\!& \frac{1}{4}\langle B_1 A_{2}B_{2}...A_rB_rA_{r+1} \rangle,\\
\no
\langle s_1^ys_{1+ r}^y \rangle  &\!=\!& ( - 1)^r\frac{1}{4}\langle A_1B_2A_2...B_rA_rB_{r+1} \rangle,\\
\langle s_1^xs_{1 + r}^y \rangle   &\!=\!& - \frac{i}{4}\langle B_1A_2B_2...A_rB_rB_{r+1} \rangle,\\
\no
\langle s_1^ys_{1 + r}^x \rangle   &\!=\!& (- 1)^r\frac{i}{4}\langle A_1B_2A_2...B_rA_rA_{r+1} \rangle, \\
\no
\langle s_1^zs_{1 + r}^z \rangle  &\!=\!& \frac{1}{4}\langle A_1B_1A_{r+1}B_{r+1} \rangle, 
\eea
%
where
%
\begin{equation} \label{eq:AB}
A_j=c_j^\dag+c_j, \ \ B_j=c_j^\dag-c_j, \ \ j=1,\ldots , r+1.
\end{equation}
%
As for the one-point function $\langle s^z_{l}\rangle$ in Eq. (\ref{eq:APB2}), one has
%
\begin{equation} \label{eq:onepoint}
\langle s^z_{l}\rangle = \frac{1}{2}\langle B_l A_l \rangle.
\end{equation}
%
 
{\color{black} The expectation values in Eq. (\ref{eq:APB3}) are defined with respect to the time-evolved fermion states that stand in for the spin states after the JW transformation. 
%
For example,
%
\begin{equation} \label{Trace0}
\langle A_1 B_2 \ldots \rangle = \mbox{Tr}(\rho(t) A_1 B_2 \ldots)
\end{equation}
%
with $\rho(t) = \otimes_k \rho_k(t)$, where $\rho_k(t)$ is the density operator for mode $k$ in a pure (mixed) ensemble of noise-free (noise-averaged) states from ramped quenches [cf. text before Eqs. (\ref{eq:amplitudes})
and (\ref{eq:AddMaster})]. Note that in the case of a pure ensemble, we used the notation $\rho_k(t) = \rho_{0,k}(t)$ in Sec. III. Also note that the product over density matrices for single modes runs over $k=\pm (2i-1)\pi/N, i=1,2,\ldots, N/2$ [cf. text after Eq. (2)].}

The transverse spin correlation functions in Eq. (\ref{eq:APB3}) can be written on the generic form
%
\begin{equation}
\langle s_1^{\alpha} s_{1+r}^{\beta} \rangle = D_r^{\alpha \beta }\langle {{\phi _1}{\phi _2}{\phi _3}...\phi _{2r-2}\phi _{2r-1}{\phi _{2r}}} \rangle,
\label{eq:APB4}
\end{equation}
%
with $D_r^{xx}= \frac{1}{4}, D_r^{yy} = ( - 1)^r\frac{1}{4}, D_r^{xy} = -\frac{i}{4},$ and $D_r^{yx} =( - 1)^r\frac{i}{4}$, where each operator $\phi_j, j= 1,2, \ldots, 2r$, 
is identified with either $A_j$ or $B_j$ according to the corresponding expression in 
Eq. (\ref{eq:APB3}). Note further that $\langle s_1^{z} s_{1+r}^{z} \rangle$ has the simple form
%
\bea
\label{eq:APB5}
\langle s_1^{z} s_{1+r}^{z} \rangle = \frac{1}{4}\langle {{\phi _1}{\phi _2}{\phi _3}{\phi_4}}\rangle,
\eea
%
where $\phi_1 = A_1, \phi_2 = B_1, \phi_3 = A_{r+1}, \phi_4 = B_{r+1}$.

The expressions in Eqs. (\ref{eq:APB4}) and (\ref{eq:APB5}) are expedient as they allow us to express the multipoint correlators in Eq (\ref{eq:APB3})
in terms of Pfaffians. This scheme was first exploited by Barouch and McCoy \cite{Barouch1971a} in their study of the
XY chain, using a result by Caianiello and Fubini \cite{caianiello} obtained via Wick's theorem.
\begin{widetext}

In this way one obtains, with $\{\alpha,\beta\} = \{x,y\}$,     
%
{\small
\bea
\label{eq:pfaffian1}
\langle s_1^\alpha s_{1+r}^\beta \rangle
&\!=\!& 
 D_r^{\alpha \beta } \mbox{pf} 
 \left( {\begin{array}{*{20}{c}}
{0}&{\langle {{\phi _1}{\phi _2}} \rangle }&{\langle {{\phi _1}{\phi _3}} \rangle }&{\langle {{\phi _1}{\phi _4}} \rangle }& \cdots &{\langle {{\phi _1}{\phi _{2r}}} \rangle }\\
{}&{0}&{\langle {{\phi _2}{\phi _3}} \rangle }&{\langle {{\phi _2}{\phi _4}} \rangle }& \cdots &{\langle {{\phi _2}{\phi _{2r}}} \rangle }\\
{}&{}&{0}&{\langle {{\phi _3}{\phi _4}} \rangle }& \cdots &{\langle {{\phi _3}{\phi _{2r}}} \rangle }\\
{}&{}&{}& \ddots & \vdots \\
{}&{}&{}&{}&{0}&{\langle {{\phi _{2r - 1}}{\phi _{2r}}} \rangle }\\
{}&{}&{}&{}&{}&{0}\\
\end{array}} \right),
\eea

\bea
\label{eq:pfaffian2}
\langle s_1^z s_{1+r}^z \rangle &=& \frac{1}{4}\mbox{pf}  \left( {\begin{array}{*{20}{c}}
{0}&{\langle {{\phi _1}{\phi _2}} \rangle }&{\langle {{\phi _1}{\phi _3}} \rangle }&{\langle {{\phi _1}{\phi _4}} \rangle }\\
{}&{0}&{\langle {{\phi _2}{\phi _3}} \rangle }&{\langle {{\phi _2}{\phi _4}} \rangle }\\
{}&{}&{0}&{\langle {{\phi _3}{\phi _4}} \rangle }\\
{}&{}&{}&{0} \\
\end{array}} \right), 
\eea
}
%

\noindent where we have written the skew-symmetric matrices on standard abbreviated form.
By combining Eqs. (\ref{eq:APB3}), (\ref{eq:APB4}), (\ref{eq:pfaffian1}), and (\ref{eq:pfaffian2}), replacing the choice $l\!=\!1$ by the collective index $l$, it is straightforward to show that the correlation functions for nearest-neighbour spins, $r\!=\!1$, and next-nearest-neighbour spins, $r\!=\!2$, are given by 
%
\bea \label{eq:APB6}
\no
\langle s^{x}_{l} s^{x}_{l+1} \rangle &=& -\frac{1}{4}\langle B_l A_{l+1}\rangle, \\ 
\no
\langle s^{y}_{l} s^{y}_{l+1} \rangle &=& -\frac{1}{4}\langle B_{l+1} A_{l}\rangle,\\
\langle s^{x}_{l} s^{y}_{l+1} \rangle &=& -\frac{i}{4}\langle B_l B_{l+1}\rangle, \\
\no
\langle s^{y}_{l} s^{x}_{l+1} \rangle &=& -\frac{i}{4}\langle A_l A_{l+1}\rangle, \\
\no
\langle s^{z}_{l} s^{z}_{l+1}\rangle &=& \frac{1}{4}\big(
\langle B_l A_{l}\rangle \langle B_{l+1} A_{l+1}\rangle
-\langle A_l A_{l+1}\rangle \langle B_{l} B_{l+1}\rangle
- \langle B_l A_{l+1}\rangle \langle B_{l+1} A_{l}\rangle \big),
\eea
%
and 
%
\bea
\label{eq:APB7}
\no
\langle s^{x}_{l} s^{x}_{l+2} \rangle & = & \frac{1}{4}\large[
\langle B_l A_{l+1}\rangle \langle B_{l+1} A_{l+2}\rangle
-\langle A_{l+1} A_{l+2}\rangle \langle B_{l} B_{l+1}\rangle
-\langle B_l A_{l+2}\rangle \langle B_{l+1} A_{l+1}\rangle \large],\\
\langle s^{y}_{l} s^{y}_{l+2} \rangle & = & \frac{1}{4}\large[
\langle B_{l+1} A_{l}\rangle \langle B_{l+2} A_{l+1}\rangle
\!-\!\langle A_{l} A_{l+1}\rangle \langle B_{l+1} B_{l+2}\rangle
-\langle B_{l+2} A_{l}\rangle \langle B_{l+2} A_{l+2}\rangle \large],\\
\no
\langle s^{x}_{l} s^{y}_{l+2} \rangle &=& -\frac{i}{4}\large[
\langle B_{l} A_{l+2}\rangle \langle B_{l+1} B_{l+2}\rangle
\!+\!\langle B_{l+2} A_{l+1}\rangle \langle B_{l} B_{l+1}\rangle
- \langle B_{l} B_{l+2}\rangle \langle B_{l+1} A_{l+2}\rangle \large],\\
\no
\langle s^{y}_{l} s^{x}_{l+2} \rangle &=& -\frac{i}{4}\large[
\langle B_{l+1} A_{l}\rangle \langle A_{l+1} A_{l+2}\rangle
\!+\!\langle B_{l+1} A_{l+2}\rangle \langle A_{l} A_{l+1}\rangle
+ \langle B_{l+1} A_{l+1}\rangle \langle A_{l} A_{l+2}\rangle
\large], \\
\no
\langle s^{z}_{l} s^{z}_{l+2}\rangle & = & \frac{1}{4}\large[
\langle B_l A_{l}\rangle \langle B_{l+2} A_{l+2}\rangle
-\langle A_{l} A_{l+2}\rangle \langle B_{l} B_{l+2}\rangle
-\langle B_{l+2} A_{l}\rangle \langle B_{l} A_{l+2}\rangle
\large],
\eea
%
respectively, with 
%
\begin{eqnarray}
\label{eq:APB8}
\no
\langle A_lA_{\l+r} \rangle &=& \langle c_l^\dag c_{\l+r}^\dag \rangle + \langle c_l c_{\l+r} \rangle + \langle c_l^\dag c_{\l+r} \rangle + \langle c_l c_{\l+r}^\dag \rangle, \\
\langle B_l B_{\l+r} \rangle &=& \langle c_l^\dag c_{\l+r}^\dag \rangle + \langle c_l c_{\l+r}\rangle - \langle c_l^\dag c_{\l+r} \rangle - \langle c_l c_{\l+r}^\dag \rangle, \\
\no
\langle {{A_l}{B_{\l+r}}} \rangle &=& - \langle B_{l+r}A_l \rangle = \langle c_l^\dag c_{\l+r}^\dag \rangle - \langle {c_l}{c_{\l+r}} \rangle - \langle c_l^\dag c_{\l+r} \rangle + \langle c_l c_{\l+r} ^\dag  \rangle, 
\end{eqnarray}
%
where, by way of translational invariance, $r=0, \pm 1, \pm 2$ cover all cases in Eqs. (\ref{eq:APB6}) and (\ref{eq:APB7}).
We obtain closed expressions for the spin correlations in these equations by calculating the fermion two-point functions in Eq. (\ref{eq:APB8}).
 Explicitly, let us consider the two-point function $\langle c_{\l}^\dagger c_{\l+r}^\dagger \rangle$, {\color{black} suppressing time arguments for
notational simplicity, and write
%
\begin{equation} \label{Trace1}
\langle c_{\l}^\dagger c_{\l+r}^\dagger \rangle = \mbox{Tr}(\rho\, c_{\l}^\dagger c_{\l+r}^\dagger),
\end{equation}
%
where, like in Eq. (\ref{Trace0}), $\rho = \otimes_k \,\rho_k$ with $\rho_k$ the density operator for mode $k$.  
By Fourier transforming, one obtains from Eq. (\ref{Trace1}),
%
\begin{eqnarray} \label{Trace2}
\langle c^\dagger_{l} c^\dagger_{l+r}\rangle &=& \frac{1}{N} \sum_{p,q} e^{-i(pl+q(l+r))} \langle  c^\dagger_p c^\dagger_q \rangle  \nonumber \\
&=& \frac{1}{N} \sum_{p,q} e^{-i(pl+q(l+r))}\mbox{Tr}(\rho \,c^\dagger_p c^\dagger_q)  \nonumber \\ 
&=& \frac{1}{N} \sum_{k_j } e^{ik_jr}\,\mbox{Tr}((\otimes_{k \neq k_j} \rho_k)\otimes\rho_{k_j}c^\dagger_{k_j}c^\dagger_{-k_j})  \\ 
&=& \frac{1}{N} \sum_{k } e^{ikr}\,\mbox{Tr}(\rho_k\, c^\dagger_k c^\dagger_{-k})  \nonumber
\end{eqnarray}
%
where, in the fourth equation, we have dropped the subindex $j$ and used $\mbox{Tr} \otimes_k \rho_k = \prod_k \mbox{Tr}\, \rho_k = 1$. The replacement of $(p,q)$ by $(k_j, -k_j)$ in the third equation follows from the fact
that each Fock space, labeled by $k_j$, is spanned by two states with zero and two quasiparticles, respectively, implying that $p=-q=k_j$ (with $k_j$ summed over) [cf. Eq. (\ref{eq:Bogostates})].

To take the trace in the last line of Eq. (\ref{Trace2}) we use the diagonal bases $\{|\phi^\pm_k\}_k$ of the mode Hamiltonians at fixed time $t$,
%
\begin{equation} \label{eq:DiagTrace}
\mbox{Tr} (\rho_kc_{k}^\dagger c_{-k}^\dagger)=\langle\phi^{-}_k|\rho_kc_{k}^\dagger c_{-k}^\dagger|\phi^{-}_k\rangle + \langle\phi^{+}_k|\rho_kc_{k}^\dagger c_{-k}^\dagger|\phi^{+}_k\rangle.
\end{equation}
%
By inserting the unit operator $\mathbb{1} = |\phi^{-}_k\rangle\langle\phi^{-}_k| + |\phi^{+}_k\rangle\langle\phi^{+}_k|$ on the right-hand side of the equation, one obtains  
%
\begin{eqnarray} 
\label{eq:cc0}
\mbox{Tr} (\rho_k c_{k}^\dagger c_{-k}^\dagger) &=& \langle\phi^{-}_k|\rho_k|\phi^{-}_k\rangle~\langle\phi^{-}_k|c_{k}^\dagger c_{-k}^\dagger|\phi^{-}_k\rangle
+ \langle\phi^{-}_k|\rho_k|\phi^{+}_k\rangle~\langle\phi^{+}_k|c_{k}^\dagger c_{-k}^\dagger|\phi^{-}_k\rangle \no \\
&+& \langle\phi^{+}_k|\rho_k|\phi^{-}_k\rangle~\langle\phi^{-}_k|c_{k}^\dagger c_{-k}^\dagger|\phi^{+}_k\rangle
+ \langle\phi^{+}_k|\rho_k|\phi^{+}_k\rangle~\langle\phi^{+}_k|c_{k}^\dagger c_{-k}^\dagger|\phi^{+}_k\rangle \no \\
& =& \rho^{(d)}_{k,11} ~\langle\phi^{-}_k|c_{k}^\dagger c_{-k}^\dagger|\phi^{-}_k\rangle + 
\rho^{(d)}_{k,12} ~\langle\phi^{+}_k|c_{k}^\dagger c_{-k}^\dagger|\phi^{-}_k\rangle\\
&+& \rho^{(d)}_{k,21}  ~\langle\phi^{-}_k|c_{k}^\dagger c_{-k}^\dagger|\phi^{+}_k\rangle 
+ \rho^{(d)}_{k,22}  ~\langle\phi^{+}_k|c_{k}^\dagger c_{-k}^\dagger|\phi^{+}_k\rangle, \no
\end{eqnarray}  
%
with the index $(d)$ a reminder that the density matrix is evaluated in the diagonal basis. The matrix elements of $c_{k}^\dagger c_{-k}^\dagger$ are easily calculated from Eq. (\ref{eq:Bogostates}),
with the result}
%
\begin{equation} \label{eq:cc1}
\langle c_{\l}^\dagger c_{\l+r}^\dagger \rangle = \frac{1}{N}\sum_{k>0} \Big[ \sin(kr) \big(\sin(2\theta_k) (\rho^{(d)}_{k,22} - \rho^{(d)}_{k,11}) + 2i (\cos^2(\theta_k) \,\rho^{(d)}_{k,12} + \sin^2(\theta_k) \,\rho^{(d)}_{k,21} )\big) \Big].
\end{equation}
%
Similarly,
%
\begin{eqnarray} 
\label{eq:cc2}
\langle c_{\l} c_{\l+r} \rangle &=& \frac{1}{N}\sum_{k>0} \Big[ \sin(kr) \big(\sin(2\theta_k) (\rho^{(d)}_{k,11} - \rho^{(d)}_{k,22}) + 2i (\cos^2(\theta_k) 
\,\rho^{(d)}_{k,21} + \sin^2(\theta_k) \,\rho^{(d)}_{k,12} )\big) \Big], \\
\label{eq:cc3}
\langle c_{\l}^\dagger c_{\l+r} \rangle &=& \frac{1}{N}\sum_{k>0} \Big[ \cos(kr) \big(2\cos^2(\theta_k) \rho^{(d)}_{k,22} + 2\sin^2(\theta_k) \rho^{(d)}_{k,11} - i\sin(2\theta_k) (\rho^{(d)}_{k,12} - \rho^{(d)}_{k,21})\big) \Big], \\
\label{eq:cc4}
\langle c_{\l} c_{\l+r}^\dagger \rangle &=& \frac{1}{N}\sum_{k>0} \Big[ \cos(kr) \big(2\cos^2(\theta_k) \rho^{(d)}_{k,11} + 2\sin^2(\theta_k) \rho^{(d)}_{k,22} - i\sin(2\theta_k) (\rho^{(d)}_{k,21} - \rho^{(d)}_{k,12})\big) \Big].
\end{eqnarray}
%
By feeding the expressions from Eqs. (\ref{eq:cc1})-(\ref{eq:cc4}) into the formulas in Eq. (\ref{eq:APB8}), we finally obtain the $A$ and $B$ two-point functions that determine the spin correlations in Eqs. (\ref{eq:APB6}) and (\ref{eq:APB7}). Pruning the outcome, the result is
%
\begin{eqnarray} \label{eq:ABfinal}
\no
\langle A_{l} A_{l+r} \rangle &=& \frac{2i}{N} \sum_{k>0} \mbox{Re}(\rho^{(d)}_{k,12})\sin(kr) + \delta_{r,0}, \\
 \langle B_{l} B_{l+r} \rangle &=& \frac{2i}{N} \sum_{k>0} \mbox{Re}(\rho^{(d)}_{k,12})\sin(kr) - \delta_{r,0}, \\
 \no
 \langle A_{l} B_{l+r} \rangle &=& - \langle B_{l+r} A_{l} \rangle, \\
 \no
&=& \frac{1}{N} \sum_{k} \Big[ (1-2\rho^{(d)}_{k,22})\big( \cos(kr)\cos(2\theta_{k}) - \sin(kr)\sin (2\theta_{k})\big)- 2\mbox{Im}(\rho^{(d)}_{k,12})\big(\cos(kr)\sin(2\theta_{k}) + \sin(kr)\cos(2\theta_{k})\big) \Big].
\end{eqnarray}
%
 
Let us mention that the two-point fermion functions above reduce to those of the equilibrium case if we take $\rho^{(d)}_{k,22}=\rho^{(d)}_{k,12}=\rho^{(d)}_{k,21}=0$:
%
\bea
\no
\langle A_{l}A_{l+r} \rangle = \delta_{r,0}, \ \ \ \langle B_{l}B_{l+r} \rangle = -\delta_{r,0}, \ \ \
\langle A_{l} B_{l+r} \rangle = -\langle B_{l+r} A_{l} \rangle = \frac{1}{N}\sum_{k}\Big[\cos(kr)\cos(2\theta_{k})-\sin(kr)\sin(2\theta_{k})\Big].
\eea
%
In the thermodynamic limit $N\rightarrow\infty$, this can be written as 
%
\bea
\no
\langle {{A_l}{A_{l+r}}} \rangle = \delta(r), \ \ \ \langle {{B_l}{B_{l+r}}} \rangle  -\delta(r), \ \ \
\langle A_{l} B_{l+r} \rangle = -\langle B_{l+r} A_{l} \rangle = \frac{1}{4\pi}\int_{0}^{2\pi} e^{ikr} e^{2i\theta_k}dk,
\eea
%
in agreement with Ref. \cite{Franchini2017}. 

Also note that for a quench from $h(t_{\text{initial}})\!\rightarrow\! -\infty$ to $h(t_{\text{final}})\!\rightarrow \!+\infty$, the off-diagonal terms $\rho^{(d)}_{k,12}(t)=(\rho^{(d)}_{k,21})^{\ast}$ of the density matrix $\rho_{k}(t)$ vanish, as does $\theta_{k}(t)$. 
As a consequence, the two-point fermion functions in Eq. (\ref{eq:ABfinal}) reduce to 
%
\bea
\no
\langle {{A_l}{A_{l+r}}} \rangle &=& \delta_{r,0}, \ \ \
\langle {{B_l}{B_{l+r}}} \rangle = -\delta_{r,0}, \\
\langle A_{l} B_{l+r} \rangle &=& -\langle B_{l+r} A_{l} \rangle = \frac{1}{N}\sum_{k} (1- 2\rho^{(d)}_{k,22})\cos(kr),
\eea
%
which in the large-time thermodynamic limit reproduces the result in Refs. \cite{Sengupta2009,Nag2011,Cherng2006}, 
%
\bea
\no
\langle {{A_l}{A_{l+r}}} \rangle &=& \delta(r), \ \ \ 
\langle {{B_l}{B_{l+r}}} \rangle = -\delta(r), \\
\langle A_{l} B_{l+r} \rangle&=&-\langle B_{l+r} A_{l} \rangle = \frac{1}{2\pi} \int_{\pi}^{\pi} (1-2p_{k}(t))\cos(kr)dk = -\frac{2}{\pi}\int_{0}^{\pi}p_{k}(t)\cos(kr)dk,
\eea
%
where $\mbox{lim}_{t\rightarrow\infty}\, \rho^{(d)}_{k,22}(t) = e^{-4\pi\sin^{2}(k)\tau} = p_{k}(t)$, with $\tau$ the time scale of the quench. 
 
\end{widetext}

Returning to our task to compute the concurrences $C_{l,l+1}(t)$ and $C_{l,l+2}(t)$ for nearest- and next-nearest-neighbor spins generated by a ramped quench, we now have all the pieces in place. From Eq. (\ref{eq:ABfinal}), we obtain closed
expressions for the spin correlations in Eqs. (\ref{eq:APB6}) and (\ref{eq:APB7}), as well as for the one-point function in Eq. (\ref{eq:onepoint}). Eq. (\ref{eq:APB2}) then yields the elements
of the reduced density matrices $\rho_{l,l+1}$ and $\rho_{l,l+2}$ [cf. Eq. (\ref{eq:APB1})]. With this, expressions for the concurrences are obtained  
from Eqs. (\ref{eq:Cdef}) and (\ref{eq:Rmatrix}). What remains is to compute the {\color{black} time-dependent matrix elements $\rho^{(d)}_{k,11}(t), \rho^{(d)}_{k,12}(t), \rho^{(d)}_{k,21}(t)$, and $\rho^{(d)}_{k,22}(t)$ that appear in Eq. (\ref{eq:ABfinal}) (with the suppressed time argument reinstated). For the noiseless pure ensemble, this can be done by numerically integrating the von Neumann equation in the diagonal basis,} 
%
\begin{equation}
\dot{\rho}_{0,k}(t)=-i[H_{0,k}(t),\rho_{0,k}(t)],
\end{equation}
%
for all $k$, with $t$ the time when the quench is completed. {\color{black} As initial condition we impose that the system is in the noiseless ground state at time $t_i$, i.e., that all modes $k$ occupy the lower levels $|\phi_k^-(t_i)\rangle$. As expected for a pure ensemble, note that the matrix elements $\rho^{(d)}_{k,11}(t), \rho^{(d)}_{k,12}(t), \rho^{(d)}_{k,21}(t)$ and $\rho^{(d)}_{k,22}(t)$ defined in Eq. (\ref{eq:cc0}) (with the time argument suppressed) map onto those in Eq. (\ref{eq:amplitudes}), where in the latter equation the density operator for mode $k$ is denoted $\rho_{0,k}(t)$ while in Eq. (\ref{eq:cc0}) the more general symbol $\rho_k$ $-$ adaptable to pure as well as mixed ensembles $-$ is used.}

In the case of a noisy ramped quench, we replace the von Neumann equation by the exact master equation \cite{luczka1991quantum,Budini2001,Chenu2017,Kiely2021}
%
\begin{equation}
\dot{\rho}_{k}(t)=-i[H_{0,k}(t),\rho_{k}(t)]- \frac{\xi^2}{2}[\sigma^z , [\sigma^z,\rho_{k}(t)]],
\label{eq:masterApp}
\end{equation}
%
where $\rho_{k}(t) = \langle \rho_{\eta,k}(t) \rangle$ is the averaged density matrix for mode $k$ in the presence of white noise, with the average taken over the white-noise distribution $\{\eta(t)\}$ specified in Eq. (\ref{eq:white}). {\color{black} By numerically integrating the master equation, Eq. (\ref{eq:masterApp}), for all $k$, with the same initial condition as in the noiseless case, we extract the matrix elements $\rho^{(d)}_{k,11}(t), \rho^{(d)}_{k,12}(t), \rho^{(d)}_{k,21}(t)$ and $\rho^{(d)}_{k,22}(t)$ in the diagonal basis. In this way we finally obtain the {\em mean concurrences} $\langle C_{l,l+1}(t) \rangle$ and $\langle C_{l,l+2}(t)\rangle$ for nearest- and next-nearest-neighbor spins. 

The computer code used to obtain the data in Figs. 1-4 can be found in Ref. \cite{WNCC}. For details on the master equation, see Appendix C.}    

\section*{Appendix C: Noise-averaged nonequilibrium concurrence: exact master equation}

\setcounter{equation}{0}
\renewcommand\theequation{C\arabic{equation}}

Let us begin by considering a general time-dependent Hamiltonian,
%
\begin{equation} \label{Hamiltonian}
H(t)=H_{0}(t)+\eta(t)H_{1}(t),
\end{equation}
%
where $H_{0}(t)$ is noise-free while $\eta(t)H_{1}(t)$ is ``noisy", with $\eta(t)$ a real-valued stochastic function for a given realization of the noise. In our application, we use white noise $\eta(t)$ with mean $\langle \eta(t)\rangle=0$ and autocorrelation function defined in Eq. (\ref{eq:white}).

To derive a master equation for the noise-averaged density matrix of $H(t)$ \cite{luczka1991quantum,Budini2001,Chenu2017,Kiely2021}, one starts by writing down the von Neumann equation
%
\begin{equation}
\label{Neumann}
\dot{\rho}_{\eta}(t)=-i[H(t),\rho_{\eta}(t)],
\end{equation}
%
where
\begin{equation} \label{quenchevolved}
\rho_{\eta}(t) = U^{\dagger}_{\eta}(t,0)\rho_{\eta}(0)U_{\eta}(t,0)
\end{equation}
is the time-evolved density matrix for a specific realization of the noise function $\eta(t)$, with $U_{\eta}(t,0) = {\cal T}\exp(-i\int_{0}^tH(t^{\prime})\, dt^{\prime})$, ${\cal T}$ being the time-ordering operator.  
 
Let $\rho(t) = \langle \rho_\eta(t) \rangle$ be the ensemble average over many noise realizations (all with a common noise-free initial condition). The averaged von Neumann equation (\ref{Neumann}) then takes the form
%
\begin{equation}
\label{meanNeumann}
\dot{\rho}(t)=-i[H_0(t),\rho(t)] - i[H_1(t),\langle \eta(t) \rho_\eta(t)\rangle].
\end{equation}
%
Exploiting a theorem by Novikov \cite{Novikov1965} yields that
%
\begin{equation}
\label{Novikov}
\langle \eta(t) \rho_\eta(t) \rangle = \langle \eta(t) \rangle \langle \rho_\eta(t) \rangle + \int_{0}^t \langle \eta(t) \eta(s) \rangle \langle \frac{\delta \rho_{\eta}}{\delta \eta} \rangle ds,
\end{equation}
%
with functional derivative
%
\begin{equation}
\label{func}
\frac{\delta \rho_{\eta}}{\delta \eta} = \frac{\partial\dot{\rho}_{\eta}}{\partial \eta} - \frac{\mbox{d}}{\mbox{d}t} \frac{\partial\dot{\rho}_{\eta}}{\partial\dot{\eta}}.
\end{equation}
%
Combining Eqs. (\ref{func}), (\ref{Hamiltonian}), and (\ref{Neumann}) gives
%
\begin{equation}
\label{func2}
\frac{\delta \rho_{\eta}}{\delta \eta} = -i[H_1(t),\eta(t)].
\end{equation}
%

The master equation follows by inserting Eq. (\ref{Novikov}) with $\langle \eta(t)\rangle=0$ into (\ref{meanNeumann}), using Eqs. (\ref{eq:white}) and (\ref{func2}),
%
\begin{equation}
\label{master}
\dot{\rho}(t) = -i[H_{0}(t),\rho(t)] - \frac{\xi^{2}}{2}\Big[H_{1}(t),[H_{1}(t),\rho(t)]\Big],
\end{equation}
%
where the factor of $1/2$ multiplying the noise term comes from the delta function in Eq. (\ref{eq:white}) having support at the edge of the integral in Eq. (\ref{Novikov}). 

As follows from the discussion in Sec. II, the TFI Hamiltonian $H_0(t)$ in Eq. (\ref{eq:Ising}) with an added white-noise term gets expressed as a sum over decoupled mode Hamiltonians $H_k(t) = H_{0,k}(t) + \eta(t)H_1$ after a Jordan-Wigner transformation, with $H_{0,k}(t)$ given in Eq. (\ref{eq:H0k}) and with $H_1=\sigma^z$. 
As a consequence, the density matrix $\rho_{\eta}(t)$ corresponding to $H_0(t)$ with the added white-noise term has a direct product structure, i.e., $\rho_{\eta}(t)=\otimes_{k}\rho_{k,\eta}(t)$ with the $2\times2$ density matrix $\rho_{k,\eta}(t)$ satisfying $\dot{\rho}_{k,\eta}(t)=-i[H_{0,k}(t),\rho_{k,\eta}(t)]$ for a single realization of the noise function $\eta$, common to all modes $k$. The master equation for the noise-averaged density matrix $\rho_{k}(t)=\langle\rho_{k,\eta}(t)\rangle$ is read off from Eq. (\ref{master}), using that $H_1=\sigma^z$, with the result
%
\begin{equation}
\label{densitymode}
\dot{\rho}_k(t)=-i[H_{0,k}(t),\rho_k(t)]-\frac{\xi^{2}}{2}\Big[\sigma^z,[\sigma^z,\rho_k(t)]\Big].
\end{equation}
%

For details on exact noise master equations, including formal properties of ensemble-averaged density matrices, we refer the reader to Refs. \cite{luczka1991quantum,Budini2001,Chenu2017,Kiely2021}.


%


\end{document}